\documentclass[journal]{IEEEtran}

\usepackage{amsmath}
\usepackage{graphicx}
\usepackage{subfigure}
\usepackage[nocompress]{cite}
\usepackage{multirow}
\usepackage{threeparttable}
\usepackage{url}
\usepackage{color}

\usepackage{balance}
\usepackage{rotating}
\usepackage{ragged2e}

\usepackage{marginnote}

\def\AmazonMicro{1-core Intel Xeon 2.5\,GHz CPU, 100-250\,Mbps}
\def\AmazonSmall{1-core Intel Xeon 2.5\,GHz CPU, 100-250\,Mbps}
\def\AmazonMedium{1-core Intel Xeon 2.5\,GHz CPU, 250\,Mbps}
\def\AmazonLarge{2-core Intel Xeon 2.5\,GHz CPU, 250\,Mbps}
\def\AmazonXLarge{4-core Intel Xeon 2.5\,GHz CPU, 1\,Gbps}
\def\AzureXSmall{1-core AMD Opteron 2.1\,GHz CPU, 30-40\,Mbps}
\def\AzureSmall{1-core Intel Xeon 2.2\,GHz CPU, 100\,Mbps}
\def\AzureMedium{2-core Intel Xeon 2.2\,GHz CPU, 200\,Mbps}
\def\AzureLarge{4-core Intel Xeon 2.2\,GHz CPU, 400\,Mbps}
\def\AzureXLarge{8-core AMD Opteron 2.1\,GHz CPU, 800\,Mbps}

\usepackage{soul} 

\def\binSize{250\,$\mu s$}

\begin{document}

\title{Fingerprinting Software-defined Networks}

\author{Heng Cui, Ghassan O. Karame,~\IEEEmembership{Member,~IEEE,} Felix
  Klaedtke, and Roberto Bifulco%
  \IEEEcompsocitemizethanks{
  \IEEEcompsocthanksitem H.~Cui,
    G.~O.~Karame, F.~Klaedtke, and R.~Bifulco are with the NEC Laboratories
    Europe, Heidelberg, Germany.
    E-mail: \mbox{firstname.lastname@neclab.eu.}}%

}
\maketitle

  \begin{abstract}
    Software-defined networking (SDN) eases network management by
    centralizing the control plane and separating it from the data
    plane. The separation of planes in SDN, however,
    introduces new vulnerabilities in SDN networks since the
    difference in processing packets at each plane allows an adversary
    to fingerprint the network's packet-forwarding logic.

    \noindent
    In this paper, we study the feasibility  of fingerprinting the controller-switch interactions
    by a remote adversary, whose aim is to acquire knowledge about
    specific flow rules that are installed at the switches.
    This knowledge empowers the adversary with a
    better understanding of the network's packet-forwarding logic
    and exposes the network to a number of threats.
    In our study, we collect measurements from  hosts located across the globe using a realistic SDN network
    comprising of OpenFlow hardware and software switches. We show that, by leveraging information from the RTT and packet-pair dispersion of the exchanged packets, fingerprinting attacks on SDN networks succeed with overwhelming probability.
    We also show that these attacks are not restricted to active
    adversaries, but can be equally mounted by passive adversaries that only monitor traffic exchanged with the SDN network.
    Finally, we discuss the implications of these attacks on the security of SDN networks, and we
    present and evaluate an efficient countermeasure to strengthen SDN networks against fingerprinting. Our results demonstrate the effectiveness of our countermeasure in deterring fingerprinting attacks on SDN networks.
  \end{abstract}


\section{Introduction}

\IEEEPARstart{S}{oftware-defined networking} (SDN)~\cite{of,nox} eases
the development and deployment of network applications by defining a
standard interface between the control plane and the data plane.  In
SDN, the control plane is implemented by a logically centralized
controller, which interacts over a bi-directional communication
channel with the data plane's network devices.  The controller can
query devices for their state, e.g., to acquire traffic statistics or
information about the status of the switches' ports, and modify their
forwarding behavior, by installing and deleting flow rules. Network
devices can also notify the controller about network events (e.g., the
reception of certain packets) and device's state changes.  For
example, a number of advanced reactive control plane logic
implementations~\cite{maple, simple, flowtags, SoftCell} configure
network devices to send notification to the controller according to
some installed policy (e.g., when a received packet does not match any
of the installed flow rules).  This notification triggers the
controller to perform a series of operations, such as installing the
appropriate forwarding rules at the switches, reserve network
resources on a given network's path, etc.

The separation of the control and data plane in SDN opens the doors
for a remote adversary to fingerprint the network. In particular, whenever packet
forwarding is performed in hardware, then packets at the data plane are
processed several orders of magnitude faster than at the
software-based control plane.
This discrepancy acts as a distinguisher
for a remote adversary to learn whether a given probe packet is
handled just at the data plane or triggers an interaction between the
data plane and the control plane. An interaction provides evidence
that the probe packet does not have any matching flow rule stored at
the switch's flow table (or it requires special attention from the
controller). This knowledge empowers an adversary with a better
understanding of the network's packet-forwarding logic and, as we
outline in this work, exposes the network to a number of threats. In
spite of the plethora of SDN security solutions in the
literature~\cite{Shin13:attacking_SDN,rosemary2014,AVANT-GUARD13,SPHINX,Hon_etal:ndss2015,Porras_etal:ndss2015},
no contribution analyzes the feasibility and realization of
fingerprinting attacks on practical SDN deployments. Moreover, there
are no proposed solutions to alleviate fingerprinting attacks on SDN.

In this paper, we address this problem and study
the fingerprinting of controller-switch interactions by a remote
adversary with respect to various network parameters, such as the
number of hops in the communication path, and the data link
bandwidth. For that purpose, we collect measurements from 20 different
hosts located across the globe (Australia, Asia, Europe, and North
America) using an SDN network comprising of several OpenFlow hardware and software
  switches. Our results show that, by leveraging information from
the packet-pair dispersion of the exchanged packets, fingerprinting
attacks on SDN networks succeed with overwhelming probability. For
instance, an adversary can correctly identify, with an
  accuracy of 98.54\%, whether a probe
  packet triggers the installation of forwarding rules at three hardware switches in our SDN network. Our results suggest that this fingerprinting
accuracy is only marginally affected by the number of hardware switches
that need to be configured on the path, and by the data link
bandwidth. More surprisingly, our results also show that the presence of
software switches, which process packets at the
software layer, does not
hinder the ability of a remote adversary in fingerprinting
controller-switch interactions.

We also show that fingerprinting attacks can be mounted by passive
adversaries that, e.g., capture a snapshot of the traffic exchanged
with the SDN network. Although existing traffic might not contain
packet pair traces, our findings show that a passive adversary can
leverage the RTT of packets (that are exchanged within a short time
interval) to fingerprint the SDN network with an accuracy up to
98.73\%. The fingerprinting accuracy due to the RTT of packets is,
however, largely affected by the SDN network size, and significantly
deteriorates with time.

This work extends our prior work in~\cite{fingerprintingSDN} (see Section~\ref{sec:related} for a detailed explanation of our extensions). To the best of our knowledge, this is the first complete work which
quantifies---by means of well defined metrics---the success of
fingerprinting attacks using state-of-the-art
OpenFlow hardware switches. We
discuss the implications of our findings on the security of SDN
networks, and we show that fingerprinting attacks can expose the SDN
network to a number of new threats that are not encountered in
traditional networks. For instance, an adversary that knows which
packets cause an interaction with the controller can, e.g., acquire
evidence about the occurrence of a particular communication event, or
abuse this knowledge to launch Denial-of-Service (DoS) attacks by
overloading the switches with bogus flow-table
updates~\cite{ofsec}. In light of our findings, we present and evaluate an efficient
countermeasure to strengthen SDN networks against fingerprinting. Our evaluation shows that our countermeasure considerably reduces the ability of an adversary to mount fingerprinting attacks on SDN networks.

The remainder of this paper is organized as follows. In
Section~\ref{sec:problem}, we define our problem statement. In
Section~\ref{sec:setup}, we describe our setup, the performed
experiments, and summarize the collected data. In
Section~\ref{sec:results}, we present and detail our results. In
Section~\ref{sec:impl}, we analyze the implications of our findings on
the security of SDN networks. In Section~\ref{sec:countermeasure}, we present and evaluate an efficient countermeasure to deter fingerprinting in SDN networks. In
Section~\ref{sec:related}, we discuss related work, and we conclude
the paper in Section~\ref{sec:conclusion}.

\section{Problem Statement}\label{sec:problem}

Before describing the focus of our work, we give a brief refresher
on OpenFlow, a widely deployed realization of SDN.

\subsection{Background}

SDN separates the control and data planes by defining a switch's programming interface and a protocol to access such interface, i.e., the OpenFlow protocol~\cite{OpenFlow_v1.3.0}.
The controller leverages the OpenFlow protocol to access the switch's programming interface and configure the forwarding behavior of the switch's data plane. The communication between the controller and switches is established using an out-of-band control channel.

The core entities exposed by the OpenFlow switch's programming
interface are flow tables and flow rules.
A flow table of a switch is just a container for its flow rules, which
define the switch's forwarding behavior.  The controller can add,
delete, or modify flow rules of a switch's flow table by sending an
$\mathtt{OFPT\_FLOW\_MOD}$ OpenFlow message to the switch.  The
parameters of an $\mathtt{OFPT\_FLOW\_MOD}$ message specify how the
flow table of the switch should be modified.
A flow rule, for instance, provides a semantic like ``if a network
packet's IP destination address is 1.2.3.4, then forward the packet to
port 2.''  In general, a flow rule contains a \textit{match set} that
defines the network packets to which the rule applies. It further
contains an \textit{action set} that defines the actions that should
be applied to such packets, for example, forward to port 2.  Whenever
a packet is received by a switch, the packet's header is used as a
search key to retrieve the rule that applies to the packet, by
performing a lookup in the flow table. The lookup operation compares
the packet's header with the rules' match set to find the rule that
\textit{matches} the packet.  Rules are prioritized in case multiple
rules match.
For the cases in which the controller needs to inspect a network
packet, before performing a forwarding decision and installing the
corresponding forwarding rules, OpenFlow defines a special ``forward
to controller'' action. When this action is applied to a packet, the
switch generates an $\mathtt{OFPT\_PACKET\_IN}$ message that is sent
to the controller. This message contains the original packet and some
additional information, such as the switch and the port ID onto which
the packet was received.

The $\mathtt{OFPT\_PACKET\_IN}$ feature is used in
  basic network control logic implementations, such as the one of an
  Ethernet learning switch. It is also used in more complex dynamic
  control plane implementations. In both cases, the network operates as follows: a packet received by the
switch generates an $\mathtt{OFPT\_PACKET\_IN}$ message; the
controller receives and analyzes the message to take a forwarding
decision; the decision is finally implemented by sending
$\mathtt{OFPT\_FLOW\_MOD}$ messages, which install rules at the
relevant switches. This ensures that all similar packets, i.e., those
that belong to the same network flow, are forwarded directly by the switches with no further interactions with the controller.
Note that the controller can use barrier messages to ensure that
message dependencies are met, e.g., when multiple switches are
reconfigured by the controller for handling a new network flow.  When
a switch receives a barrier request
($\mathtt{OFPT\_BARRIER\_REQUEST}$), the switch must finish all
previously received messages before processing new messages. After
processing all messages, it notifies the controller by sending a
barrier reply message ($\mathtt{OFPT\_BARRIER\_REPLY}$).

\subsection{Problem Statement}

The main objective of our work is to study the ability of a remote
adversary to identify whether an interaction between the controller
and the switches (and a subsequent rule installation) has been triggered by a given packet.
The absence of a controller-switch interaction typically provides evidence that the flow rules that handle the received packet are already installed at the switches. Otherwise, if a communication between the controller and the switches is triggered, then this suggests that the received packet requires further examination by the controller, e.g., since it does not have any matching entry stored at the switch's flow table, or because the controller requires additional information before installing a forwarding decision at the switches.

In our study, we consider both active and passive adversaries. We assume that an active adversary can compromise a remote client, inject probe packets of her choice, and capture
the timing of the corresponding responses issued by a
server. In contrast, a passive adversary cannot inject packets in the network but only monitors the exchanged traffic between the server and the client.
Notice that passive adversaries are hard to detect by standard intrusion detection systems since they do not generate any extra network traffic.

Our study focuses on answering the following questions:
\begin{itemize}
\item Is it possible to remotely identify whether the installation of flow rules has been triggered by a given packet?
\item What is the accuracy of fingerprinting attacks in SDN networks?
\item What is the impact of the number of switches that need to be configured on the
  fingerprinting accuracy?
\item What is the impact of the data link bandwidth on the fingerprinting accuracy?
\item Is the fingerprinting accuracy affected by the presence of software switches in the SDN network?
\item How and to which extent can such fingerprinting attacks be efficiently mitigated?
\end{itemize}

\section{Experimental Setup}\label{sec:setup}

In this section, we detail our experimental setup. This includes a
description of our testbed, the used features, the conducted experiments,
and the collected datasets.

\subsection{Testbed}
\label{subsec:setup}

\begin{figure*}[tb]
  \centering
  \includegraphics[scale=0.08]{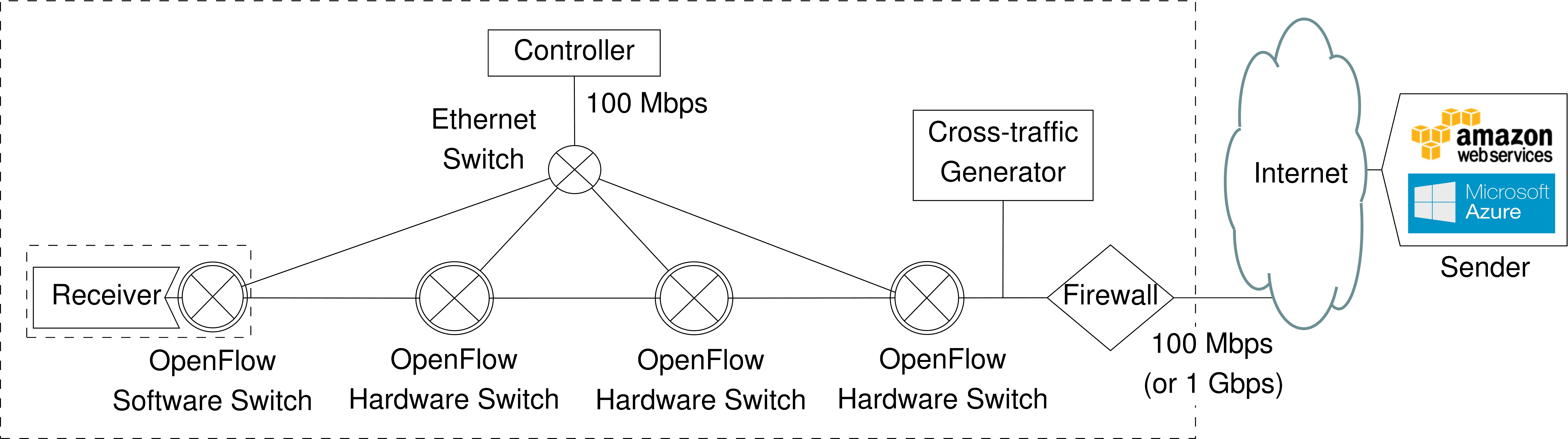}
  \caption{Sketch of our measurement setup. Our testbed comprises three
    NEC PF5240 OpenFlow hardware switches and one OpenVSwitch
    (version 2.3.1).}
  \label{fig:exp_topo}
\end{figure*}
Our measurement setup is summarized in Figure~\ref{fig:exp_topo}.  The
testbed comprises three OpenFlow hardware switches (three NEC PF5240
switches~\cite{necsw}) and one OpenFlow software switch (an
OpenVSwitch, version 2.3.1~\cite{Pfaff_etal:OpenVSwitch}).  The
switches are connected to the data plane over a 100\,Mbps data
channel. We also consider the case where the data channel bandwidth
increases to 1\,Gbps.  Note that, although our testbed only comprises
three hardware switches, it can emulate the processing of packets in
many realistic datacenters. Recall that a conventional datacenter's
network typically consists of three layers of switches: top-of-rack,
aggregation, and core~\cite{datacenter}. Packets are usually processed
by at most one switch in each of these layers, that is, each packet
traverses (at most) three hops in the datacenter's network.  The
testbed's switches interface with a Floodlight v0.9
controller~\cite{floodlight}, which runs on a computer with a 6-core
Intel Xeon L5640 2.26\,GHz CPU and 24\,GB of RAM.  A legacy Ethernet
switch bridges the connections between the OpenFlow switches and the
controller. To emulate realistic network load on the control channel,
we limit the control interface of the switches to 100\,Mbps.

The controller is configured to minimize the processing delay for an incoming \textit{packet-in} event, i.e., we only require the controller to perform a table lookup and retrieve pre-computed forwarding rules in response to \textit{packet-in} events.
Furthermore, the controller always performs bi-directional flow installation; that is, the handling of a \textit{packet-in} event
triggers the installation of a pair of rules, one per flow direction,
at each involved switch. We ensure that the controller's CPU is not
overloaded during our measurements.

We deploy a cross-traffic generator on
an AMD dual core processor running at 2.5\,GHz
to emulate realistic WAN traffic load on the switches' ports that were
used in our study. The generated
cross traffic follows
a Pareto distribution with 20\,ms mean and 4\,ms
variance~\cite{Dobre13:PoWerStore}.

To analyze the effect of the data link bandwidth on the fingerprinting
accuracy, we bridge our SDN network to the Internet using 100\,Mbps
and 1\,Gbps links (respectively), by means of a firewall running on an
AMD Athlon dual core processor 3800+ machine. For the purpose of our
experiments, we collect measurement traces between an Intel Xeon
E3-1230 3.20\,GHz CPU server with 16\,GB RAM and 20 remote clients
deployed across the globe. Table~\ref{tab:exp_instance_type} details
the specifications and locations of the clients used in our
experiments. In our testbed, the server and the software
switch were co-located on the same machine.

Note that, by reducing the time required for rule installation to a
minimum, our testbed emulates a scenario that is particularly hard for
fingerprinting. In Section~\ref{subsec:impl}, we discuss the
implications of our setup on our findings.

\subsection{Features}\label{sec:feature}

We define the network path $P_{\mathit{cs}}$ between a client $c$ and a
server $s$ as a sequence of consecutive links $P_{\mathit{cs}} =\langle L_{\mathit{cs}}^1,
\dots,L_{\mathit{cs}}^n\rangle$, which are connected via network components. Similarly, we denote the reverse path $P_{sc}$ by $P_{\mathit{sc}}
 =\langle L_{\mathit{sc}}^1,\dots,L_{\mathit{sc}}^m\rangle$.
We denote by $\tau_{L^i}$ the transmission delay at hop $i$; that is, $\tau_{L^i} =
\frac{S}{B_{L^i}}$, where $S$ is the size of the packet and $B_{L^i}$
is the capacity of $L^{i}$. Furthermore, let $d_{L^i}^j$ refer to the
additional delay that is experienced by packet $j$ when traversing
$L^{i}$. The delay $d_{L^i}^j$ generally results from
additional queuing exhibited by packet $j$ due to cross-traffic at hop
$i$.

To identify a communication event between the switches and the
controller, we rely on two time-based features: packet-pair dispersion
and RTT.

\subsubsection{Packet-pair Dispersion}

The \emph{dispersion} between two packets sent by the client after a
link $L_{\mathit{cs}}^{i}$ refers to the time interval between the
complete transmission of these packets on $L_{\mathit{cs}}^{i}$. When
measuring the dispersion of a packet pair traversing an SDN network,
two cases emerge.

\vspace{1 em}\paragraph*{\normalfont\textbf{Case 1---Packets do not trigger rule installation}}
Assuming that two probe packets (labeled in the sequel by ``1'' and ``2'') are sent by the client with an initial dispersion $\Delta_{0}$, and that they do not trigger any interaction on the control plane, then the resulting
dispersion measured after a link $L_{\mathit{cs}}^{i}$ is given
by~\cite{DispersionTechniques, Karame13:SecDisp}:
\begin{equation}
  \Delta_{i} =
  \begin{cases}
    \tau_{L_{\mathit{cs}}^i} + {d}_{L_{\mathit{cs}}^i}^2
    & \text{if $\tau_{L_{\mathit{cs}}^i} + {d}_{L_{\mathit{cs}}^i}^1 \geq \Delta_{(i-1)}$}\\
    \Delta_{(i-1)} + ({d}_{L_{\mathit{cs}}^i}^2 - {d}_{L_{\mathit{cs}}^i}^1)
    & \text{otherwise.}
  \end{cases}
  \label{eq:packet}
\end{equation}

In our setup, the client sends \emph{large} packet pairs back-to-back in time with an initial
dispersion $\Delta_{0} = \frac{S}{B_{L_{\mathit{cs}}^0}}$. These packets are
then highly likely to queue at the bottleneck link (the link for which $B^{\mathit{cs}}_i$ is
  minimal). Let $\mathit{min}$ be the index of the bottleneck link on
  the internet path $P_{\mathit{cs}}$. Following from Equation~\ref{eq:packet},
$\Delta_{n}$ (measured by the server) is then given by:
\begin{equation*}
 \Delta_{n} = \frac{S}{B^{\mathit{cs}}_{\mathit{min}}} +
 {d}_{L_{\mathit{cs}}^{\mathit{min}}}^2 +
 \sum_{i=\mathit{min}+1}^{n}({d}_{L_{\mathit{cs}}^i}^2-{d}_{L_{\mathit{cs}}^i}^1)
 \,,
\end{equation*}
where ${d}_{L_{\mathit{cs}}^{\mathit{min}}}^2$ refers to the additional queuing delay that is
experienced by the second packet on the bottleneck link.
Notice that in the absence of cross-traffic, $\Delta_{n}\approx
\frac{S}{B^{\mathit{cs}}_{\mathit{min}}}$. If the server immediately
issues small replies to the packets sent by the client (e.g., by
issuing ACKs), then these packets are unlikely to queue on the reverse path $P_{\mathit{sc}}$, and the measured dispersion between the reply packets will approximately correspond to $\Delta_{n}$~\cite{DispersionTechniques}. As shown in~\cite{DispersionTechniques, Karame13:SecDisp}, the packet-pair dispersion is a feature that is relatively stable over time (since it depends on the bottleneck bandwidth of the path).

\vspace{1 em}\paragraph*{\normalfont\textbf{Case 2---Packets trigger rule installation}}
$\Delta_{n}$ is typically in the order of tens of microseconds in
current Internet paths~\cite{secure_bandwidth_2009}. However, we
expect $\Delta_{n}$ to increase (e.g., to the order of few
milliseconds) if the probe packet pair triggered an interaction on the
control plane. This is mainly due to the (relatively slow) handling of
a notification by the controller.\footnote{Before any packet is
  forwarded by the switch, it undergoes the following steps:
  \emph{(i)} the packet (or just its header) is transmitted to the
  controller; \emph{(ii)} the controller performs a table-lookup in
  order to invoke the corresponding forwarding rule for the packet;
  \emph{(iii)} the decision is transmitted to the involved switch(es)
  in the form of a flow table entry; \emph{(iv)} the switch installs
  the entry, and, finally, the packet is forwarded by the
  switch.} Namely, when the packet pair triggers an interaction on
the control plane, then:
\begin{equation*}
 \Delta_{n} = \frac{S}{B^{\mathit{cs}}_{\mathit{min}}} +
 {d}_{L_{\mathit{cs}}^{\mathit{min}}}^2 +
 \sum_{i=\mathit{min}+1}^{n}({d}_{L_{\mathit{cs}}^i}^2-{d}_{L_{\mathit{cs}}^i}^1)
 + \max_{\forall k}\delta^i_{k}
 \,.
\end{equation*}
Here, $\delta^{i}_k$ refers to the delay introduced by a possible
communication between the controller and OpenFlow switch $k$ on the path between the sender and receiver. Since we assume that the controller installs bi-directional rules on all switches at once, only the \emph{maximum} installation delay is accounted (and is only witnessed
when packets traverse $P_{\mathit{cs}}$).

\begin{table*}[tb]
  \centering
  \caption{Remote clients used in our experiments. Bandwidths
    are based on estimates from the cloud providers.}
  \scalebox{1.0}{
  \begin{threeparttable}
  \begin{tabular}{|@{\ }c@{\ }|c|c|}
    \hline &  Location  &  Profile Details \\
    \hline\hline
    \multirow{10}{*}{\begin{turn}{90}Amazon\end{turn}} & \multirow{5}{*}{Europe} & \AmazonMicro \\ \cline{3-3}
    &   &  \AmazonSmall \\ \cline{3-3}
    &   &  \AmazonMedium   \\ \cline{3-3}
    &   &  \AmazonLarge    \\ \cline{3-3}
    &   &  \AmazonXLarge  \\ \cline{2-3}
    & \multirow{2}{*}{Australia} & \AmazonSmall \\ \cline{3-3}
    &   &  \AmazonLarge   \\ \cline{2-3}
    & \multirow{3}{*}{North America} & \AmazonMicro  \\ \cline{3-3}
    &   &  \AmazonMedium \\ \cline{3-3}
    &   &  \AmazonXLarge \\ \cline{3-3}
    \hline
    \hline
    \multirow{10}{*}{\begin{turn}{90}Azure\end{turn}} &
    \multirow{5}{*}{Europe} & \AzureXSmall\tnote{$\dagger$} \\ \cline{3-3}
    &   &  \AzureMedium  \\ \cline{3-3}
    &   &  \AzureXLarge \\ \cline{3-3}
    &   &  \AzureSmall  \\ \cline{3-3}
    &   &  \AzureLarge \\ \cline{2-3}
    & \multirow{2}{*}{Asia} & \AzureSmall  \\ \cline{3-3}
    &   &  \AzureLarge  \\ \cline{2-3}
    & \multirow{3}{*}{North America} & \AzureXSmall\tnote{$\dagger$} \\ \cline{3-3}
    &   &  \AzureMedium \\ \cline{3-3}
    &   &  \AzureXLarge  \\ \cline{3-3}
    \hline
  \end{tabular}
  \begin{tablenotes}
  \item[$\dagger$] These bandwidths were obtained using measurements.
  \end{tablenotes}
  \end{threeparttable}}
  \label{tab:exp_instance_type}
\end{table*}

\subsubsection{Round Trip Times (RTT)}

The RTT witnessed by a packet $i$ sent from the client to the server
is:
\begin{align}\label{eq:rtt}
\mathit{RTT}_i =  \sum_{j=1}^{n} (\tau_{L_{\mathit{cs}}^j} +
{d}_{L_{\mathit{cs}}^j}^i) + \sum_{j=1}^{m} (\tau_{L_{sc}^j} +
{d}_{L_{\mathit{sc}}^j}^i) + \max_{\forall k}\delta^i_{k}
\,.
\end{align}
Clearly, if no communication between switch $k$ and the controller occurs (e.g., forwarding rules are already installed), then $\delta^{i}_k =0$. Since there might be more than one OpenFlow switch on $P_{\mathit{cs}}$,
$\mathit{RTT}_i$ depends on the maximum latency incurred by a switch-controller interaction across all the OpenFlow switches included in $P_{\mathit{cs}}$.

Since the RTT exhibited by packets largely depends on the geographical
location of hosts, and on the underlying network condition, we measure
in our experiments the difference, $\delta_{\mathit{RTT}}$, between the RTT of
two probe packets issued by the same sender, i.e., $\delta_{\mathit{RTT}}
  =\mathit{RTT}_1-\mathit{RTT}_2$. This feature does not
depend on the location of hosts, but is mainly dominated by rule installation overhead and network jitter. Namely, following from
Equation~\ref{eq:rtt}:
\begin{align*}
\delta_{\mathit{RTT}} &= \max_{\forall k}\delta^1_{k} -\max_{\forall
  k}\delta^2_{k} + \sum_{j=1}^{n} {d}_{L_{\mathit{cs}}^j}^i
+\sum_{j=1}^{m} {d}_{L_{\mathit{sc}}^j}^i
\,,
\end{align*}
where $\sum_{j=1}^{n} {d}_{L_{\mathit{cs}}^j}^i +\sum_{j=1}^{m} {d}_{L_{\mathit{sc}}^j}^i$ is typically negligible.
If both packets do not result in
any rule installation, then $\delta_{\mathit{RTT}}\approx
0$. Otherwise, if one of the packets
triggers a rule installation,
then $|\delta_{\mathit{RTT}}| \gg 0$, since
$\max_{\forall k}\delta^1_{k}\gg 0$ or $\max_{\forall k}\delta^2_{k} \gg 0$.

\subsection{Data Collection}\label{subsec:data}

To collect timing information based on our features, we deployed 20 remote clients across the globe (cf. Table~\ref{tab:exp_instance_type}) that exchange UDP-based probe packet trains
with the local server. Notice that we rely on UDP for transmitting packets
  since Internet gateways may filter TCP SYN or ICMP
  packets.

Each probe train consists of:
\begin{itemize}
\item A $\mathtt{CLEAR}$ packet signaling the start of the measurements. Upon reception of this packet, the controller deletes all the entries stored within the flow tables of the OpenFlow switches in $P_{\mathit{cs}}$.
\item After one second\footnote{Our experimental results show that one
    second is enough to account for rule installation on all
    four OpenFlow switches in our network.} since the transmission of the
  $\mathtt{CLEAR}$ packet, the client transmits four MTU-sized packet pairs. Here, different packet pairs are sent with an additional second of separation.
\item After one second since the transmission of the last packet pair, another $\mathtt{CLEAR}$ packet is sent to clear all flow tables.
\item Two packets separated by one second finally close the probe train.
\end{itemize}

We point out that all of our probe packets belong to the same network
flow, i.e., they are crafted with the same packet header. For each
received packet of every train, the local server issues a short reply
(e.g., a 64 bytes ACK). We maintain a detailed log of the timing information relevant to the sending and reception of the exchanged probe packets. When measuring dispersion, we account for out-of-order packets; this explains negative dispersion values.

For each of our 20 clients, we exchange 450 probe trains on the paths
$P_{\mathit{cs}}$ and $P_{\mathit{sc}}$ to the server. Half of these
probe trains are exchanged before noon, while the remaining half is
exchanged in the evening. In our measurements, we vary the number of
OpenFlow switches that need to be configured in reaction to the
exchanged probe packets. Namely, we consider the following four cases
where a probe packets triggers the reconfiguration of some of the
OpenFlow switches: (1)~one hardware switch, (2)~two hardware switches,
(3)~three hardware switches, and (4)~the software switch. We remark
that the choice of the configured hardware switches in our testbed
(cf. Figure~\ref{fig:exp_topo}) has no impact on the measured features
since we ensure that the remaining hardware switches have already
matching rules installed.  Furthermore, we remark that packets of a
probe train only traverse the software switch in case~(4), i.e., when
it is configured.  In total, our data collection phase lapsed from
April 27, 2015 until October 27,2015, in which 869,201 probe packets
were exchanged with our local server using all clients/configurations,
amounting to almost 0.66\,GB of data.

\subsection{Evaluation Metric}

We evaluate two hypotheses based on our features: \emph{(i)} the first
hypothesis states that no rule installation was triggered by our probe
packets and \emph{(ii)} the second hypothesis corresponds to the
conjecture that a rule was installed in reaction to our probes.  Here,
there are two possible errors: \emph{false match} and \emph{false
  non-match}. In our case, the former is equivalent to a decision
that no rule was installed, while in reality our probes triggered the
installation of a rule.  The latter is equivalent to a decision that a
rule was installed, while in reality no rule was installed.  The
\emph{False Match Rate} (FMR) and \emph{False Non-match Rate} (FNR)
represent the frequencies at which these errors occur. The \emph{Equal
  Error Rate} (EER), which is used as a single metric for the accuracy
of an identification system~\cite{fvc}, is the rate at which FMR and
FNR are equal. In the sequel, we use the EER to evaluate the
effectiveness of our features.

We compute the EER as follows. We compute the \emph{Probability
  Distribution Function} (PDF) of the measured values of our features
(across all configurations and clients location) as described in
Sections~\ref{subsec:setup} and~\ref{sec:feature}. We then separate
the PDFs in two categories: \emph{(i)} $\mathit{PDF}_{N}$ that
contains all measurements obtained when our probes did not trigger a
rule installation, and \emph{(ii)} $\mathit{PDF}_{Y}$ that contains
those measurements obtained when the probe packets caused a rule
installation at $k$ OpenFlow switches (with $k=1,2,3$ hardware
switches or $k=1$ software switch). We then compute the rate of
falsely accepted and falsely rejected hypotheses given a threshold.
The measurements from $\mathit{PDF}_{N}$ that are above this threshold
indicate the number of false rejects (FNR), and measurements from
$\mathit{PDF}_{Y}$ that are below the threshold indicate the number of
false accepts (FMR). Recall that the EER is the error rate where FNR
and FMR are equal. The value of the EER-based threshold is our
reference for an accept/reject decision. If the value of a measurement
is smaller than the threshold, then we conjecture it belongs to
$\mathit{PDF}_{N}$; otherwise, we conjecture that it belongs to
$\mathit{PDF}_{Y}$.

Note that EER values are between $0\%$ and $100\%$. An EER value for a feature
close to $50\%$ indicates that our hypotheses cannot be distinguished
from each other for the given feature. In particular, the value $50\%$
means that $\mathit{PDF}_{N}$ and $\mathit{PDF}_{Y}$ for
the given feature completely overlap, and, based on the feature, an
adversary cannot distinguish at all whether a packet triggered a rule
installation.  Conversely, EER values close to $0\%$ and $100\%$
indicate that our hypotheses are distinguishable based on our
features, i.e., the fingerprinting accuracy is high.

\section{Evaluation Results}\label{sec:results}

In this section, we present and analyze our experimental results using each of our proposed time-based features.

\subsection{Packet-pair Dispersion Feature}

\begin{figure*}[tb]
  \centering
  \subfigure[$k=3$ hardware switches]{\label{fig:disp_3sw}\includegraphics[width=0.245 \textwidth]{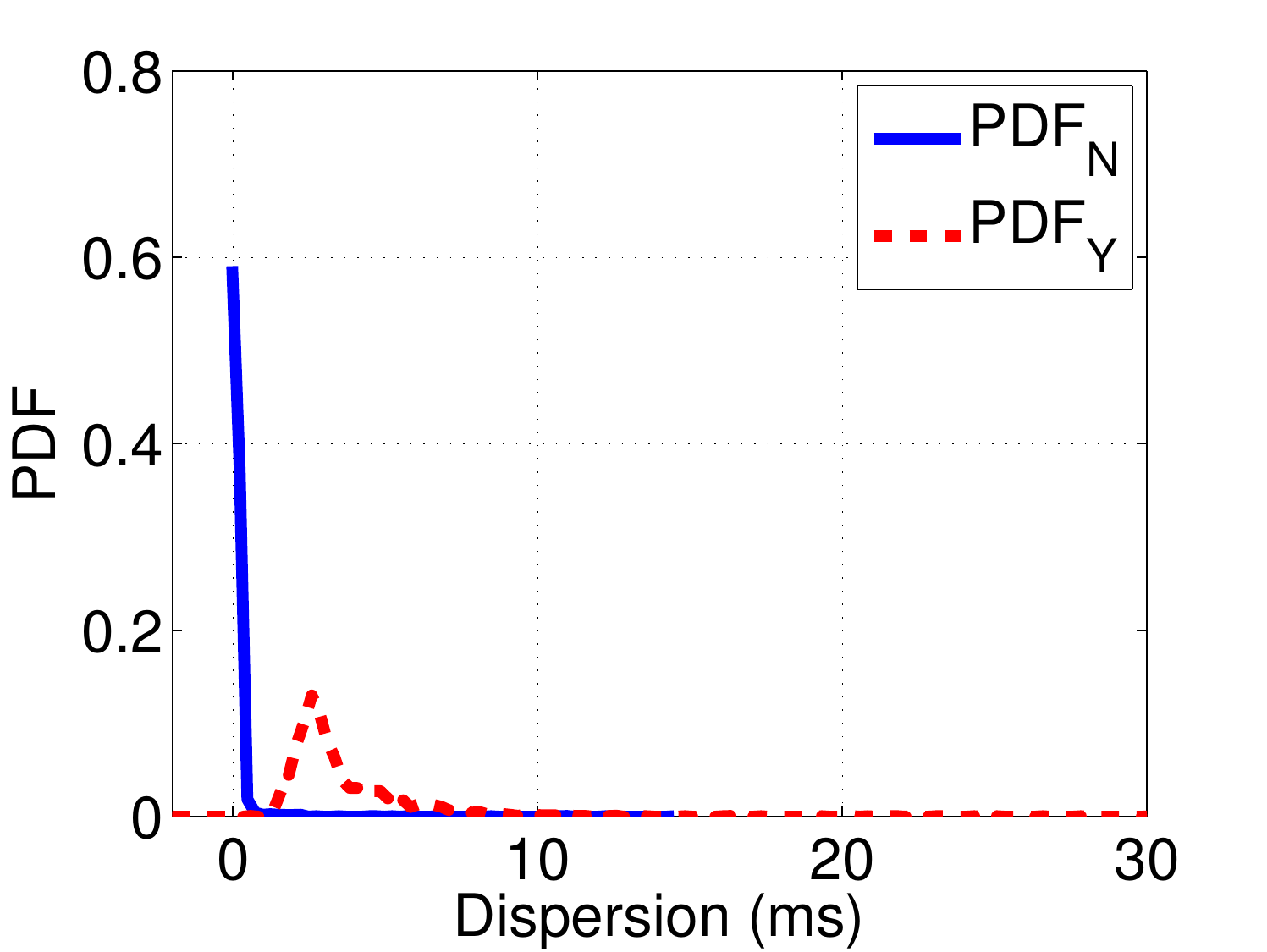}}
  \subfigure[$k=2$ hardware switches]{\label{fig:disp_2sw}\includegraphics[width=0.245 \textwidth]{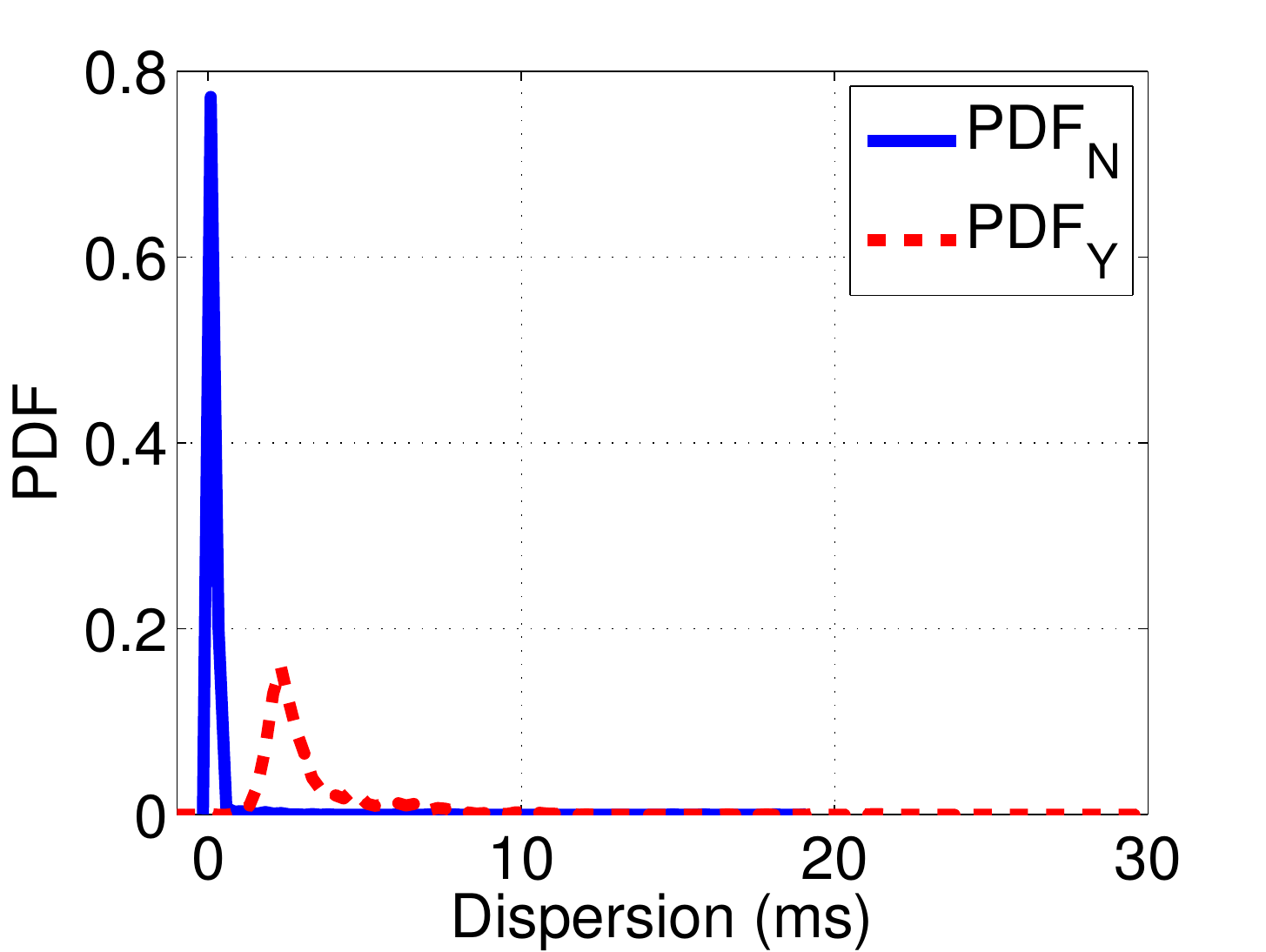}}
  \subfigure[$k=1$ hardware switch]{\label{fig:disp_1sw}\includegraphics[width=0.245 \textwidth]{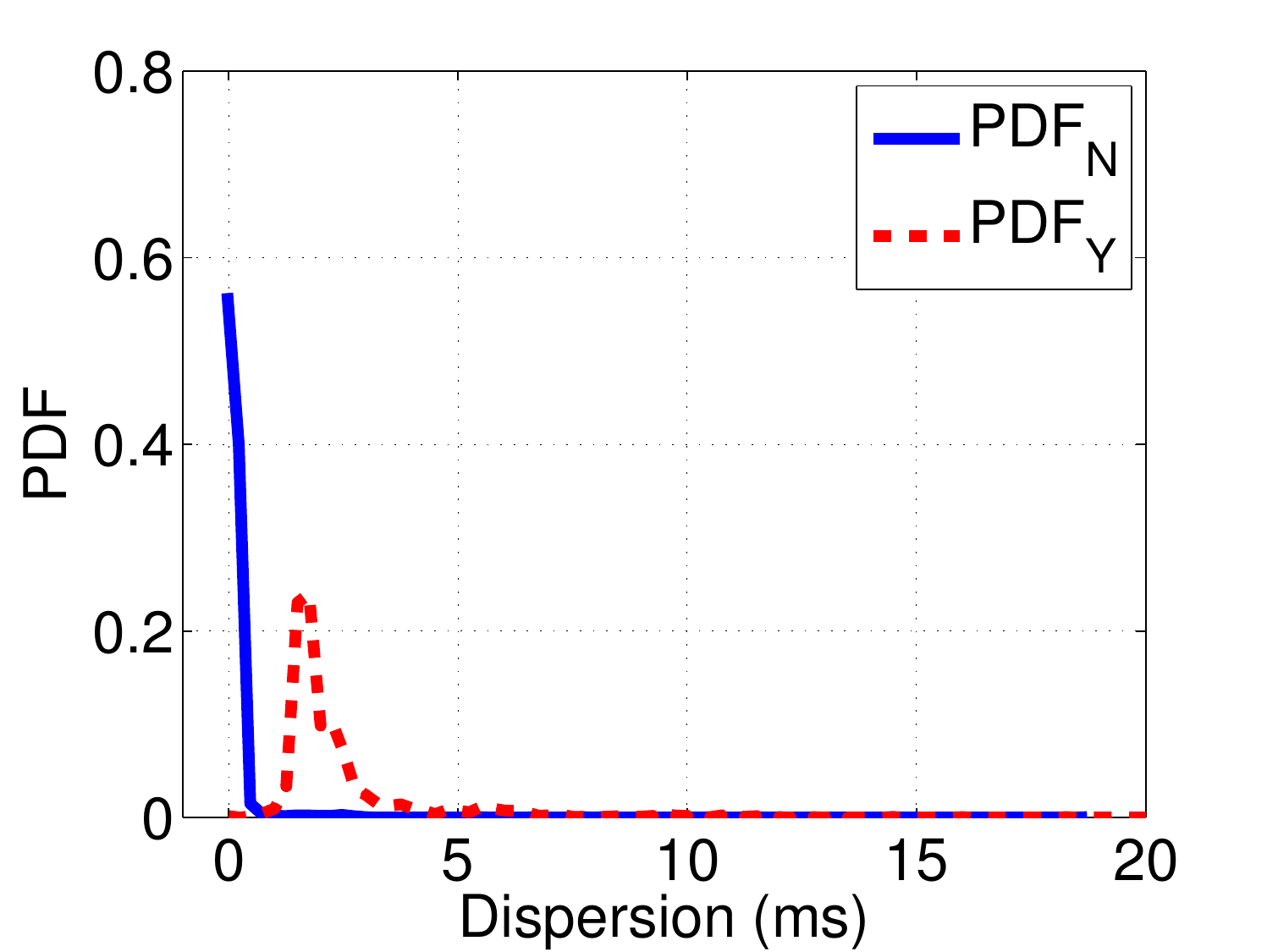}}
  \subfigure[$k=1$ software switch]{\label{fig:disp_1soft}\includegraphics[width=0.245 \textwidth]{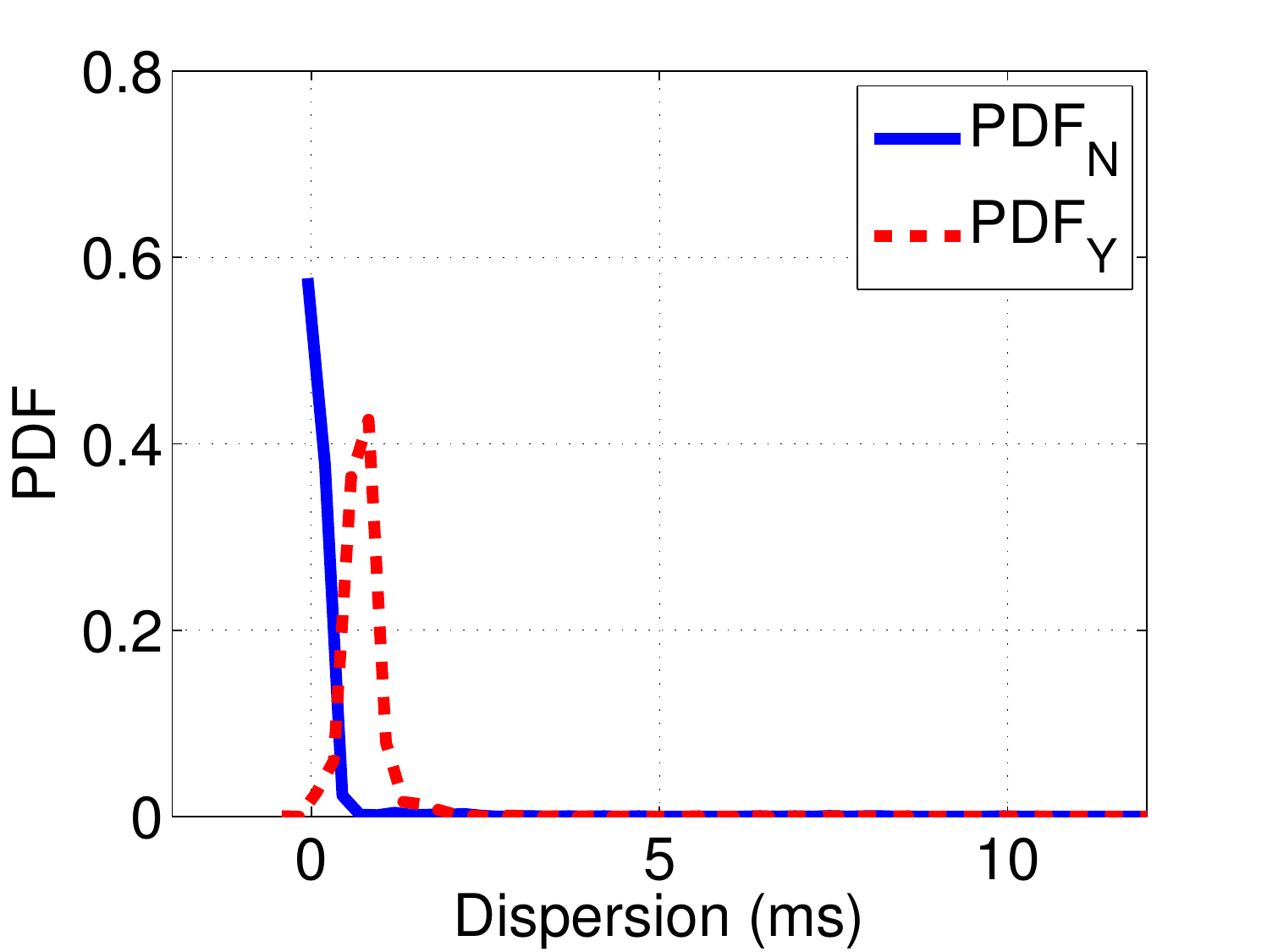}}
  \caption{Fingerprinting SDN networks (100\,Mbps link) using
    packet-pair dispersions. In our plots, we assume a bin size of
    \binSize.}
  \label{fig:pdf_disp}
\end{figure*}
\begin{figure*}[tb]
  \centering
  \subfigure[$k=3$ hardware switches]{\label{fig:disp_3hw_1gbps}\includegraphics[width=0.245 \textwidth]{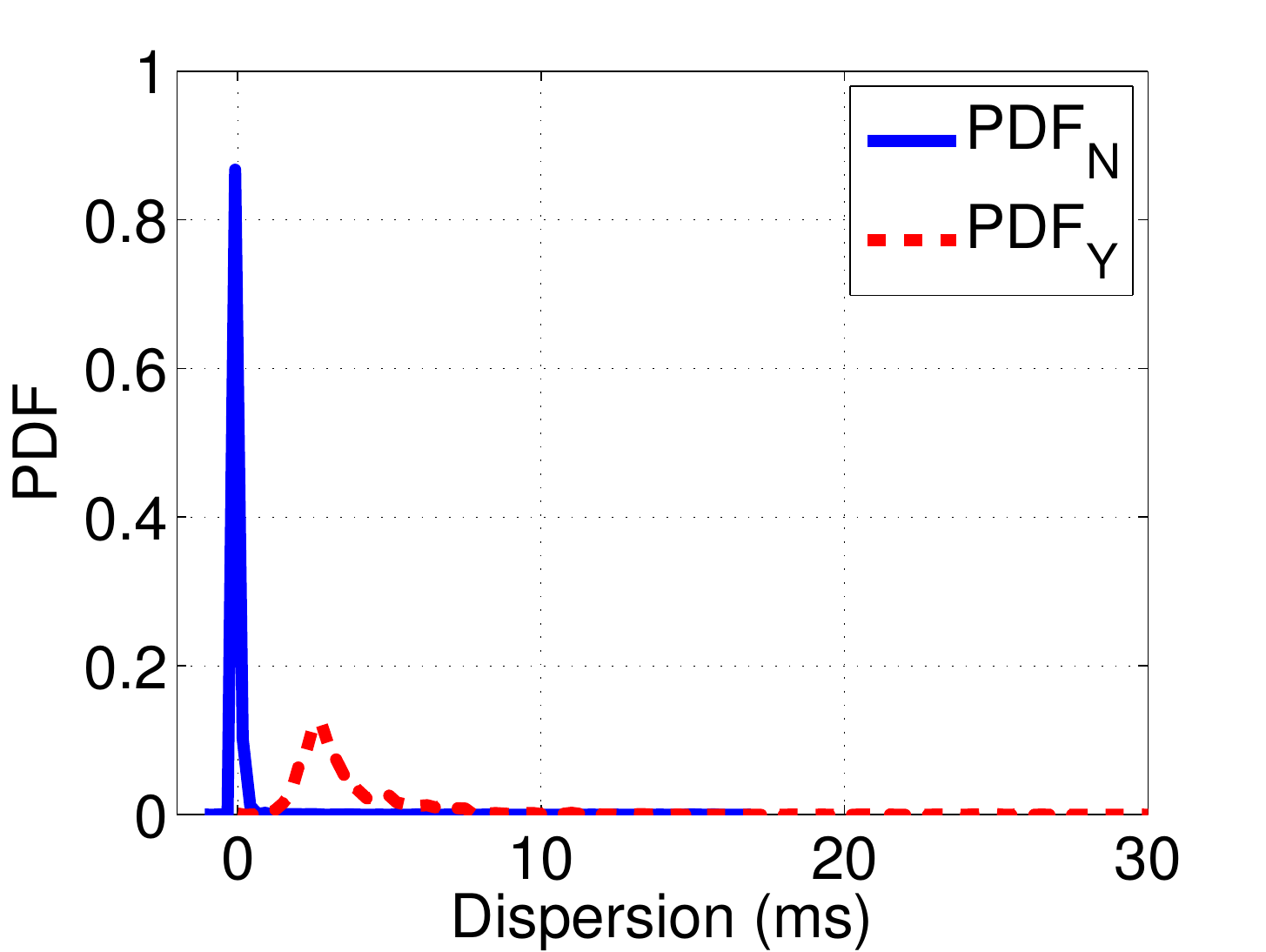}}
  \subfigure[$k=2$ hardware switches]{\label{fig:disp_2hw_1gbps}\includegraphics[width=0.245 \textwidth]{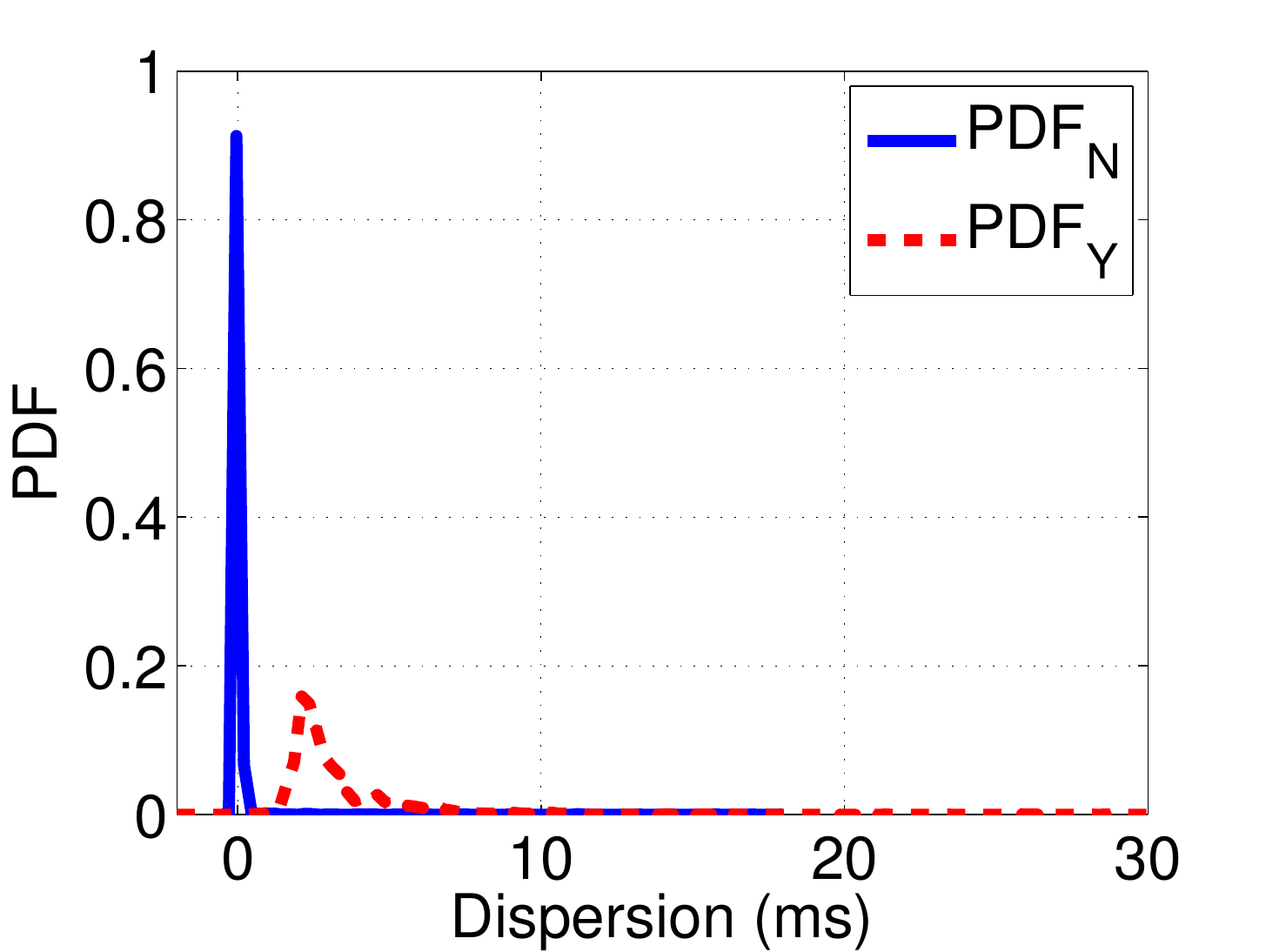}}
  \subfigure[$k=1$ hardware switch]{\label{fig:disp_1hw_1gbps}\includegraphics[width=0.245 \textwidth]{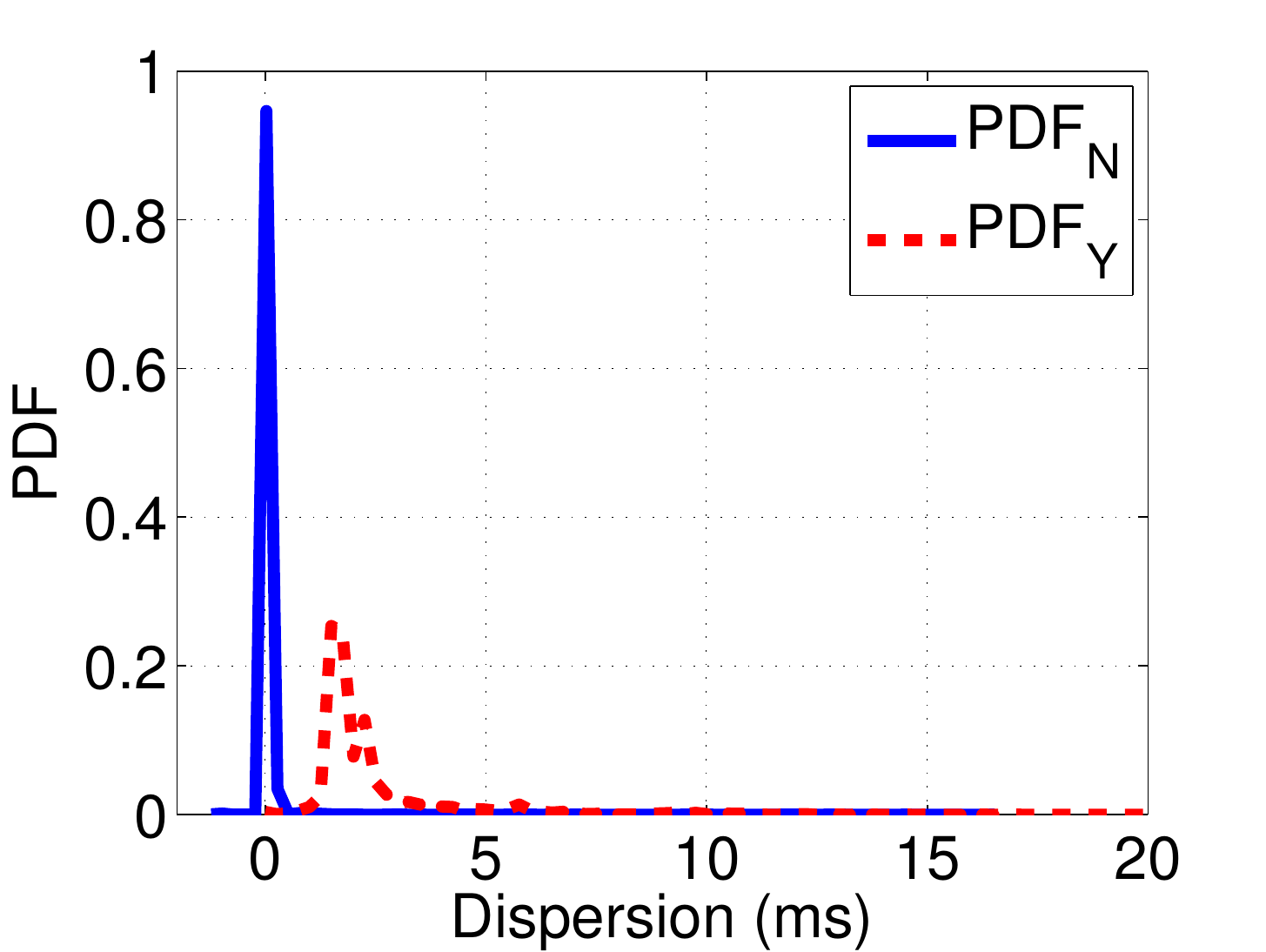}}
  \subfigure[$k=1$ software switch]{\label{fig:disp_1soft_1gbps}\includegraphics[width=0.245 \textwidth]{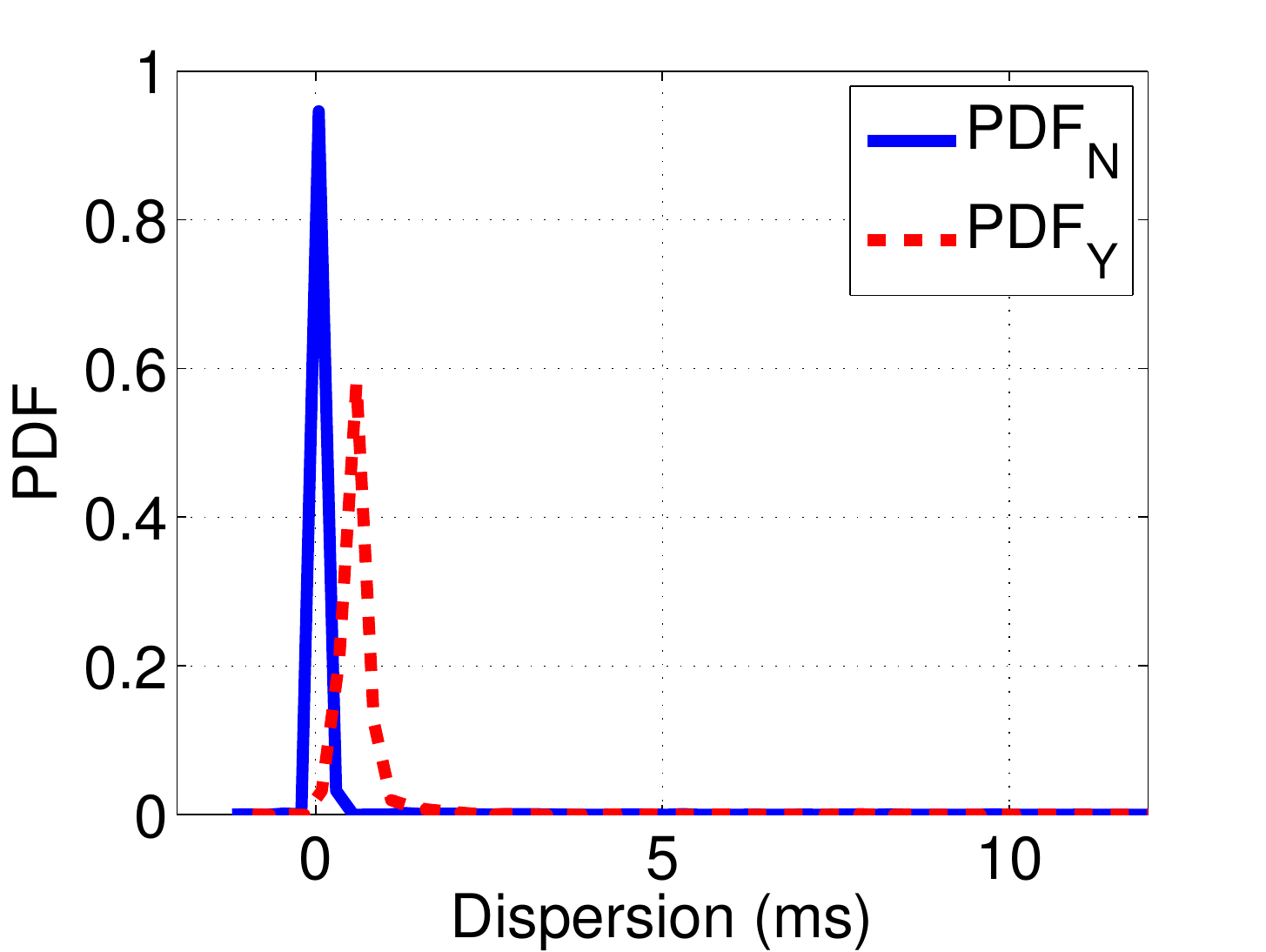}}
  \caption{Fingerprinting SDN networks (1\,Gbps link) using
    packet-pair dispersions. In our plots, we assume a bin size of
    \binSize.}
  \label{fig:pdf_disp_1gbps}
\end{figure*}
In Figure~\ref{fig:pdf_disp}, we show \emph{(i)} the PDF of dispersion
values for which none of the packets of a pair
triggered any rule installation at the switches (referred to as
$\mathit{PDF}_{N}$), and \emph{(ii)} the PDF of dispersion values for which the
probes triggered a rule installation (referred to as
$\mathit{PDF}_Y$).

Our results show that the sample mean of $\mathit{PDF}_Y$ is considerably
greater than that of $\mathit{PDF}_N$; $\mathit{PDF}_Y$ and $\mathit{PDF}_N$ are significantly
different at 1\% according to t-test~\cite{ttest} when the number $k$
of OpenFlow hardware and software switches that need to be configured varies between 1 and 3. This is mainly due to the fact that the delay required for rule installation, $\max_{\forall k}\delta^i_{k}$, acts as a strong distinguisher when measuring $\Delta_n$.
More specifically, our results show that across all locations the
obtained EER are approximately 1\% for $k=2,3$ hardware switches,
1.74\% for $k=1$ hardware switch, and 4.49\% for
$k=1$ software switch. For example, when $k=3$ hardware switches, the EER is calculated using a threshold of 1.43\,ms;
that is, 98.92\% of all measured values in $\mathit{PDF}_N$ are below 1.43\,ms, and 98.92\% of all measured values above 1.43\,ms are contained in $\mathit{PDF}_Y$.

As shown in Figure~\ref{fig:pdf_disp_1gbps}, our results are negligibly
affected by the data link bandwidth. Notably, the EER marginally
increases by almost 0.2\% for $k=2,3$ hardware switches when the bandwidth of the data link increases
from 100\,Mbps to 1\,Gbps. In this setting, the EER decreases by 0.5\%
when $k=1$ hardware or software switch.

\subsection{RTT Feature}

\begin{figure*}[tb]
  \centering
  \subcapraggedrighttrue
  \subcaphangtrue
  \subfigure[$k=3$ hardware switches, time~span~1\,second]{\label{fig:delta_rtt_3sw}\includegraphics[width=0.245\textwidth]{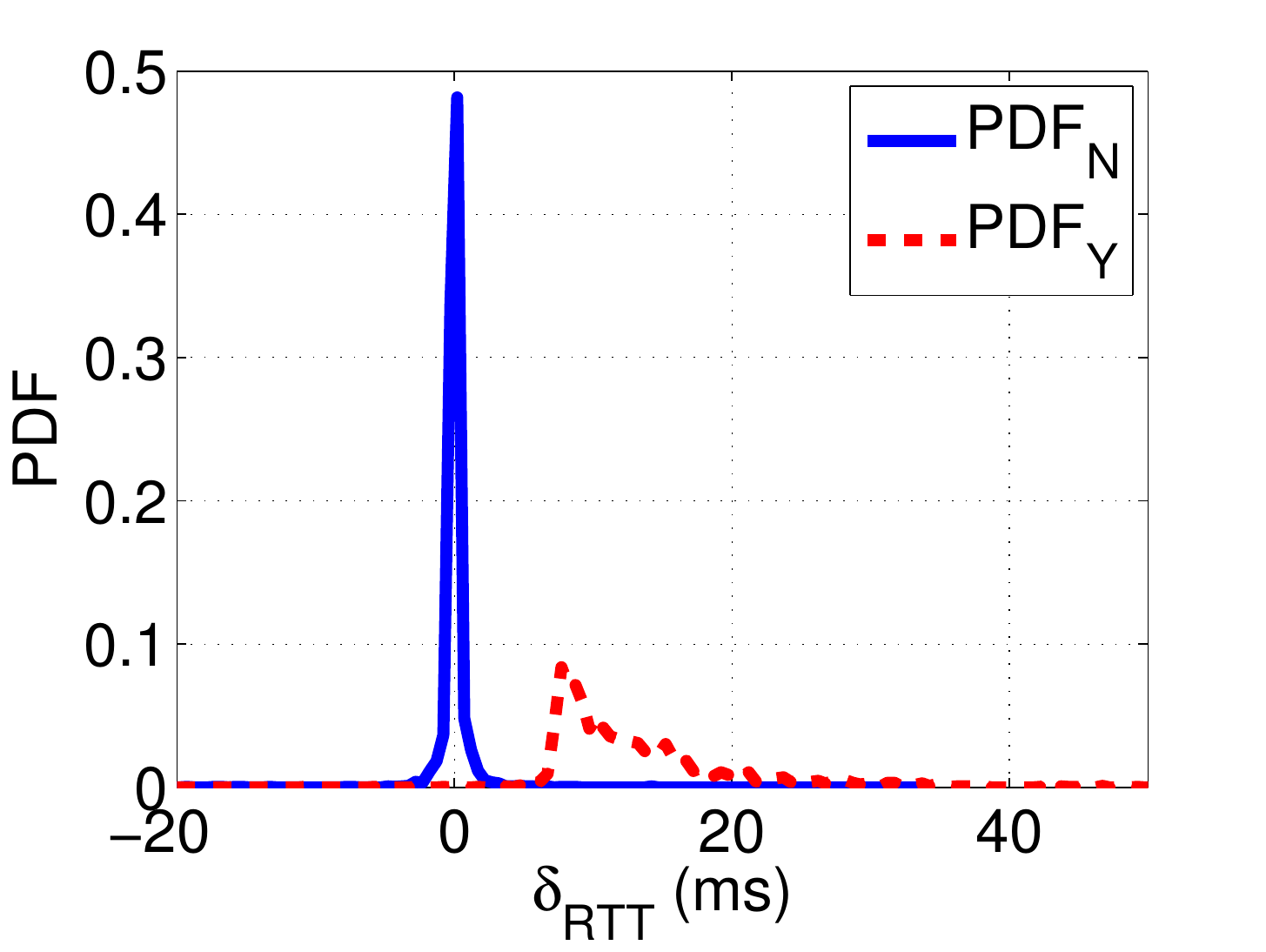}}
  \subfigure[$k=2$ hardware switches, time~span~1\,second]{\label{fig:delta_rtt_2sw}\includegraphics[width=0.245\textwidth]{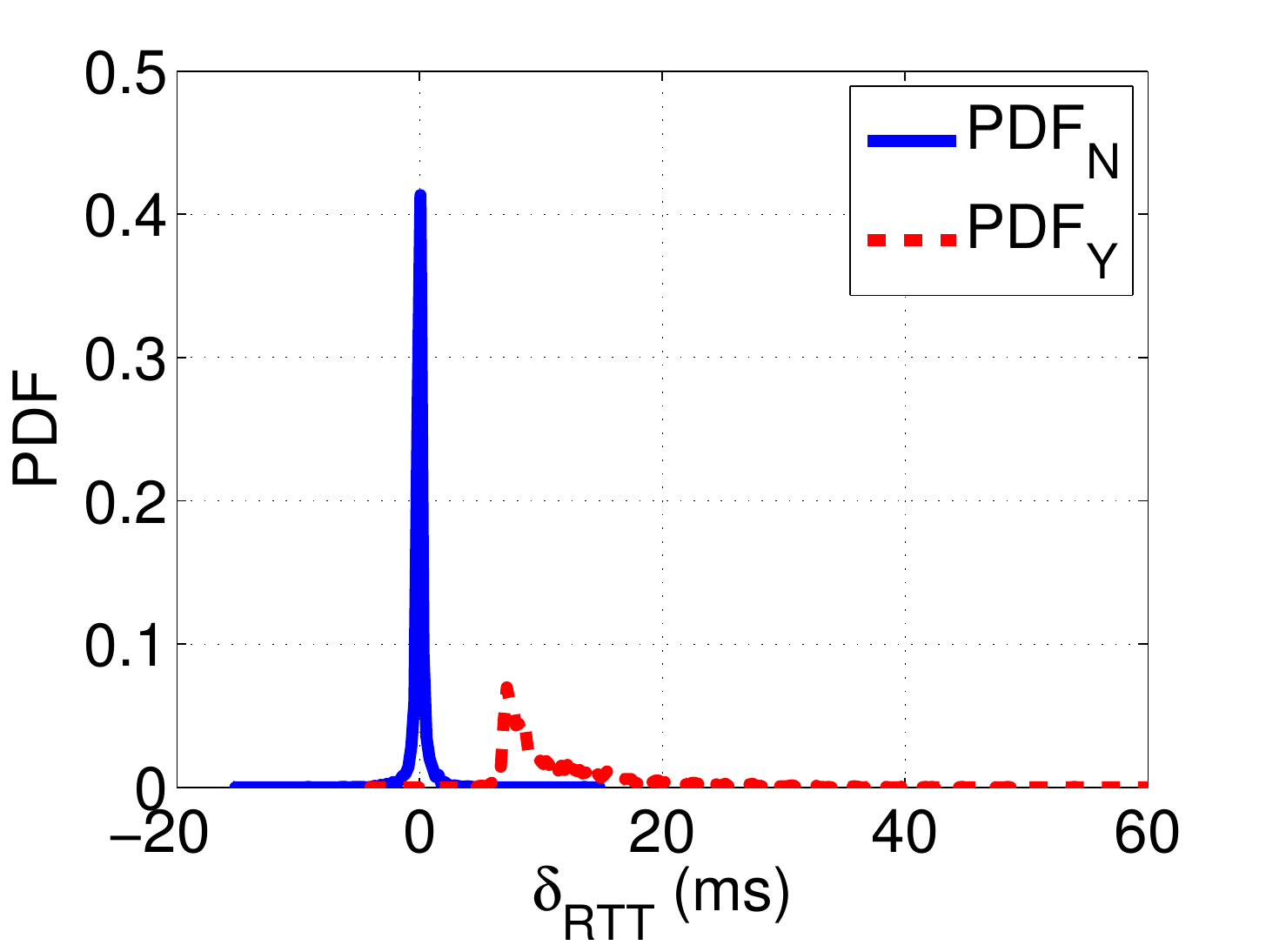}}
  \subfigure[$k=1$ hardware switch, time~span~1\,second]{\label{fig:delta_rtt_1sw}\includegraphics[width=0.245\textwidth]{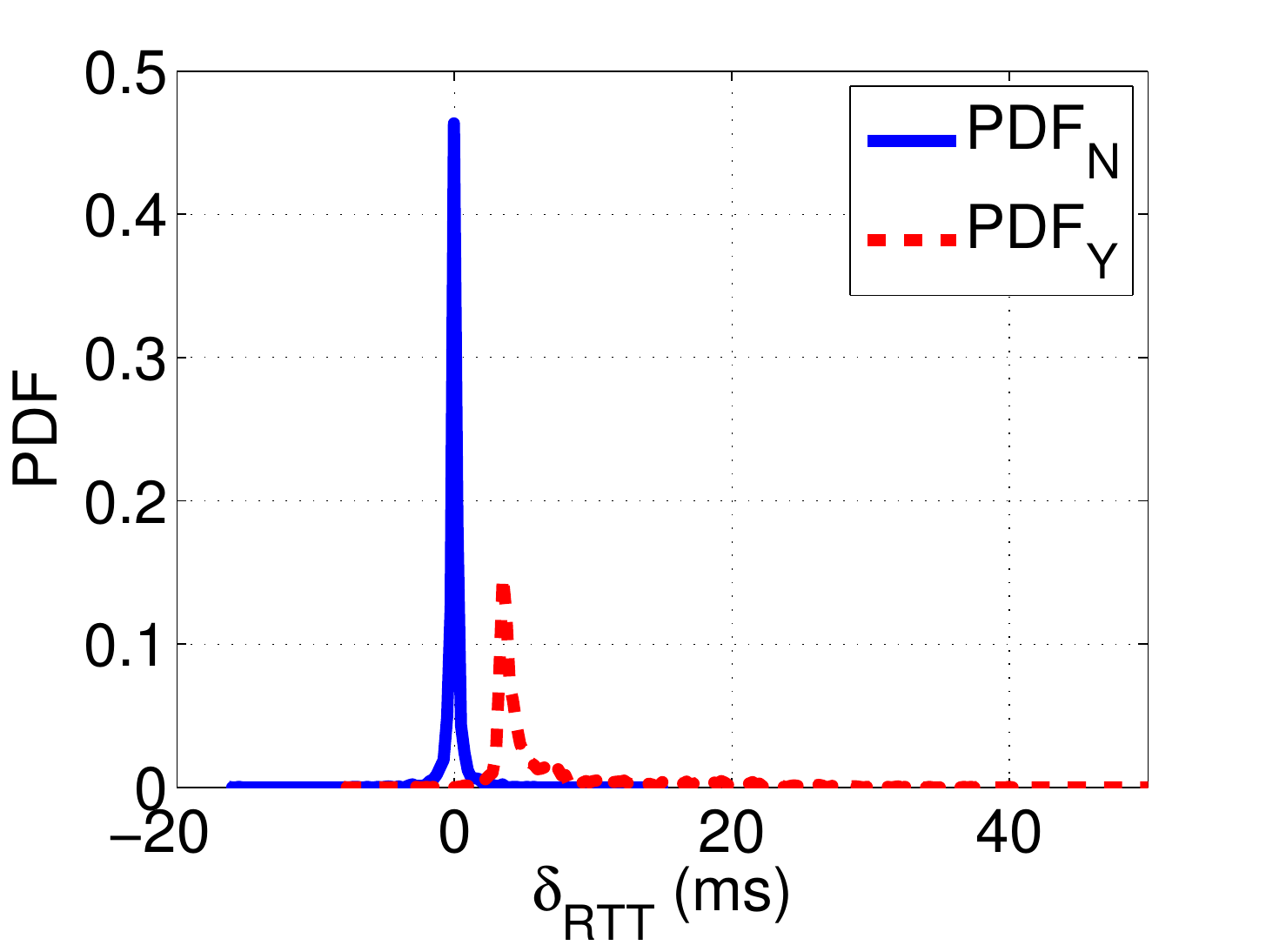}}
  \subfigure[$k=1$ software switch, time~span~1\,second]{\label{fig:delta_rtt_1soft}\includegraphics[width=0.245\textwidth]{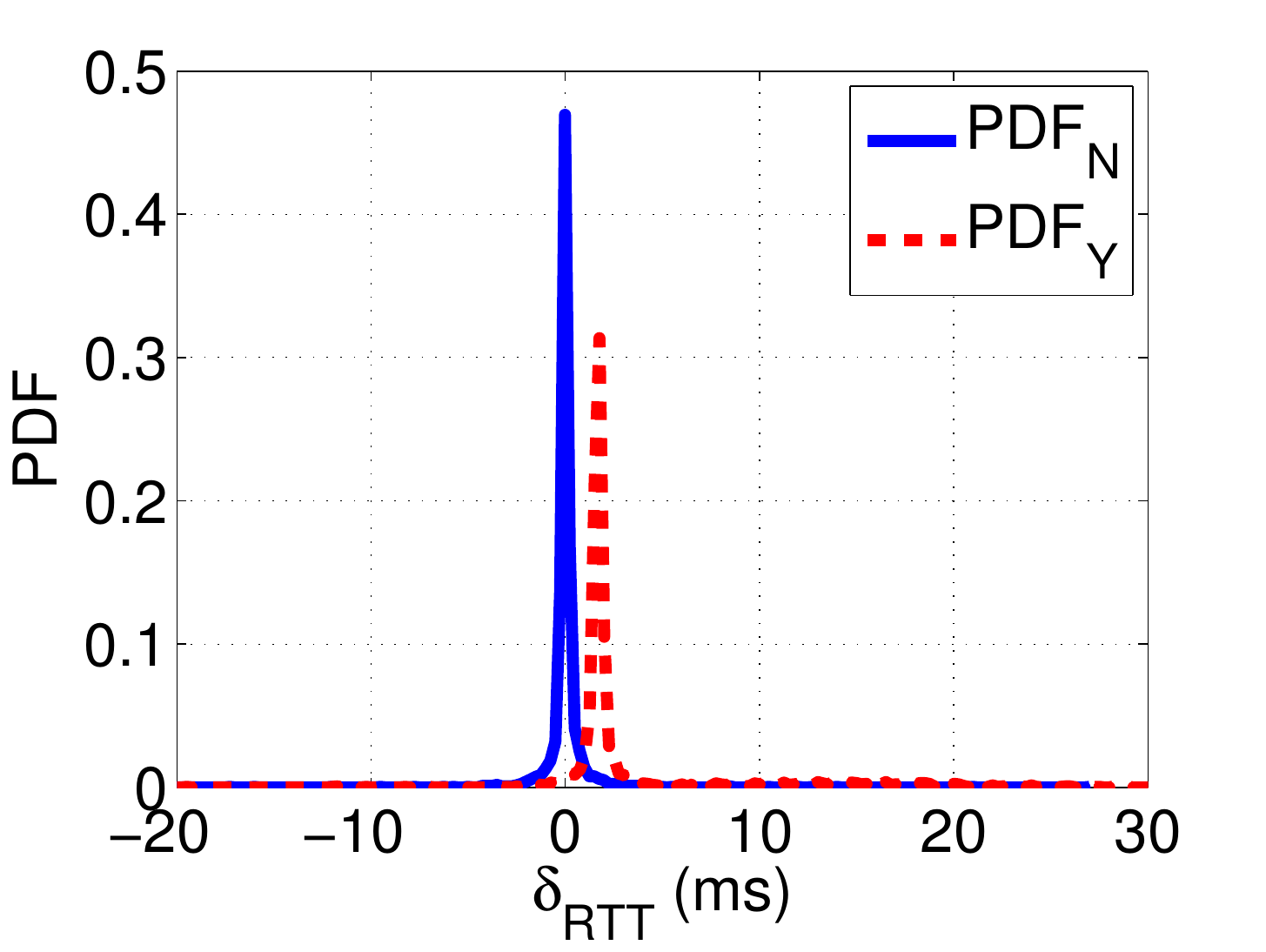}}
  \caption{Fingerprinting SDN networks (100\,Mbps link) using
    $\delta_{\mathit{RTT}}$. In our plots, we assume a bin size of
    \binSize.}
  \label{fig:pdf_rtt}
\end{figure*}
\begin{figure*}[tb]
  \centering
  \subcapraggedrighttrue
  \subcaphangtrue
  \subfigure[$k=3$ hardware switches, time~span~1\,second]{\label{fig:delta_rtt_3hw_1gbps}\includegraphics[width=0.245\textwidth]{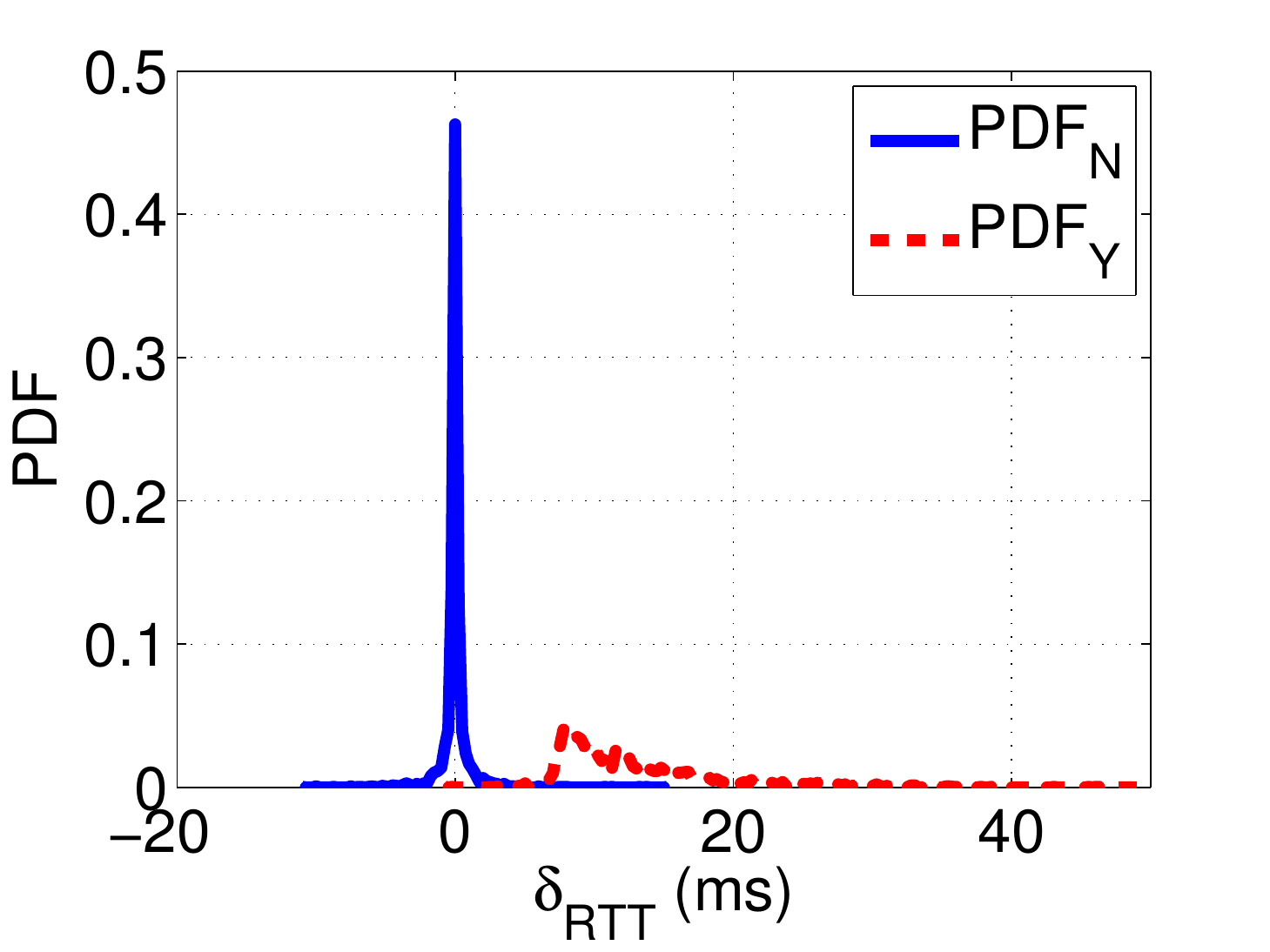}}
  \subfigure[$k=2$ hardware switches, time~span~1\,second]{\label{fig:delta_rtt_2hw_1gbps}\includegraphics[width=0.245\textwidth]{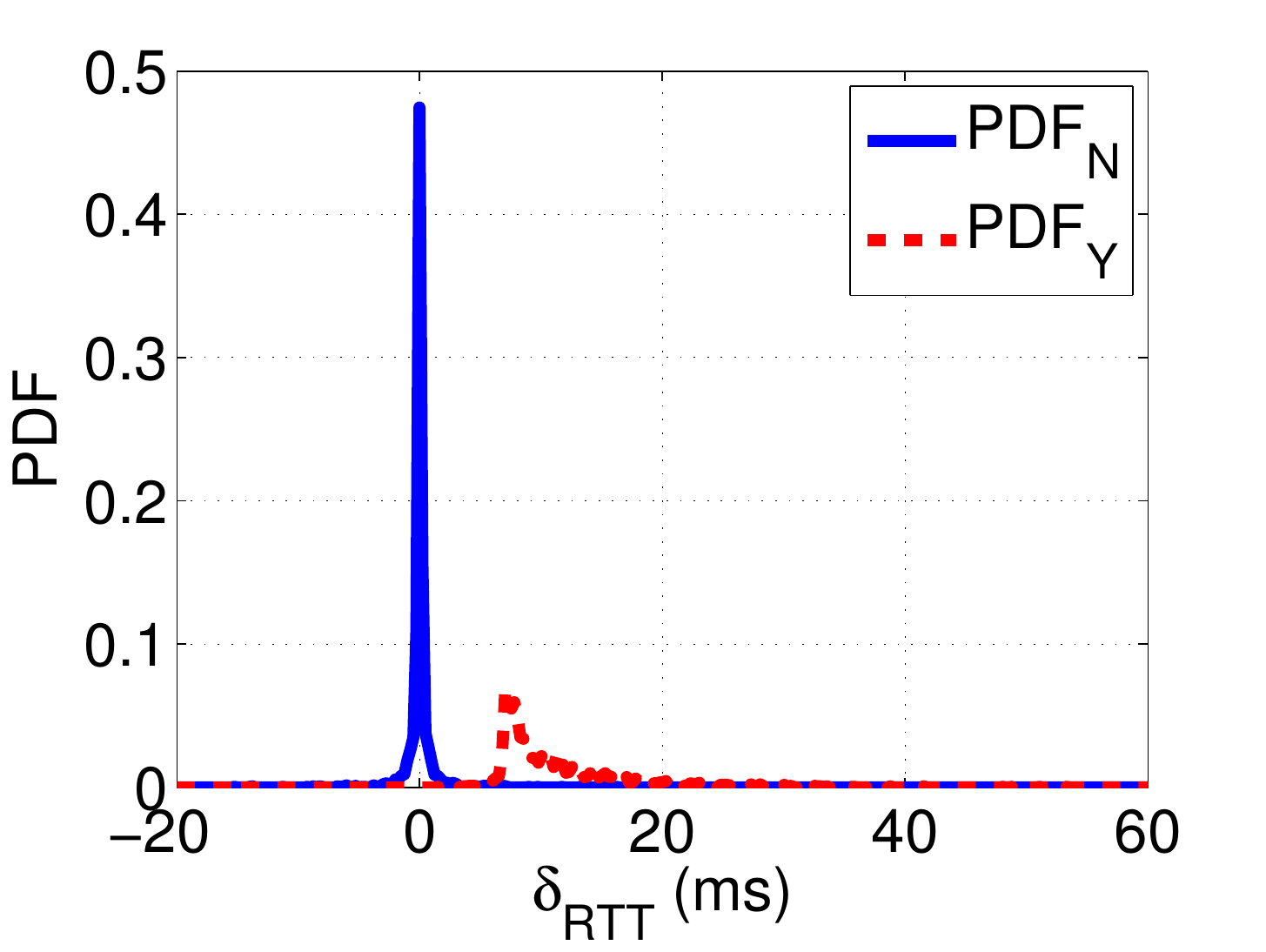}}
  \subfigure[$k=1$ hardware switch, time~span~1\,second]{\label{fig:delta_rtt_1hw_1gbps}\includegraphics[width=0.245\textwidth]{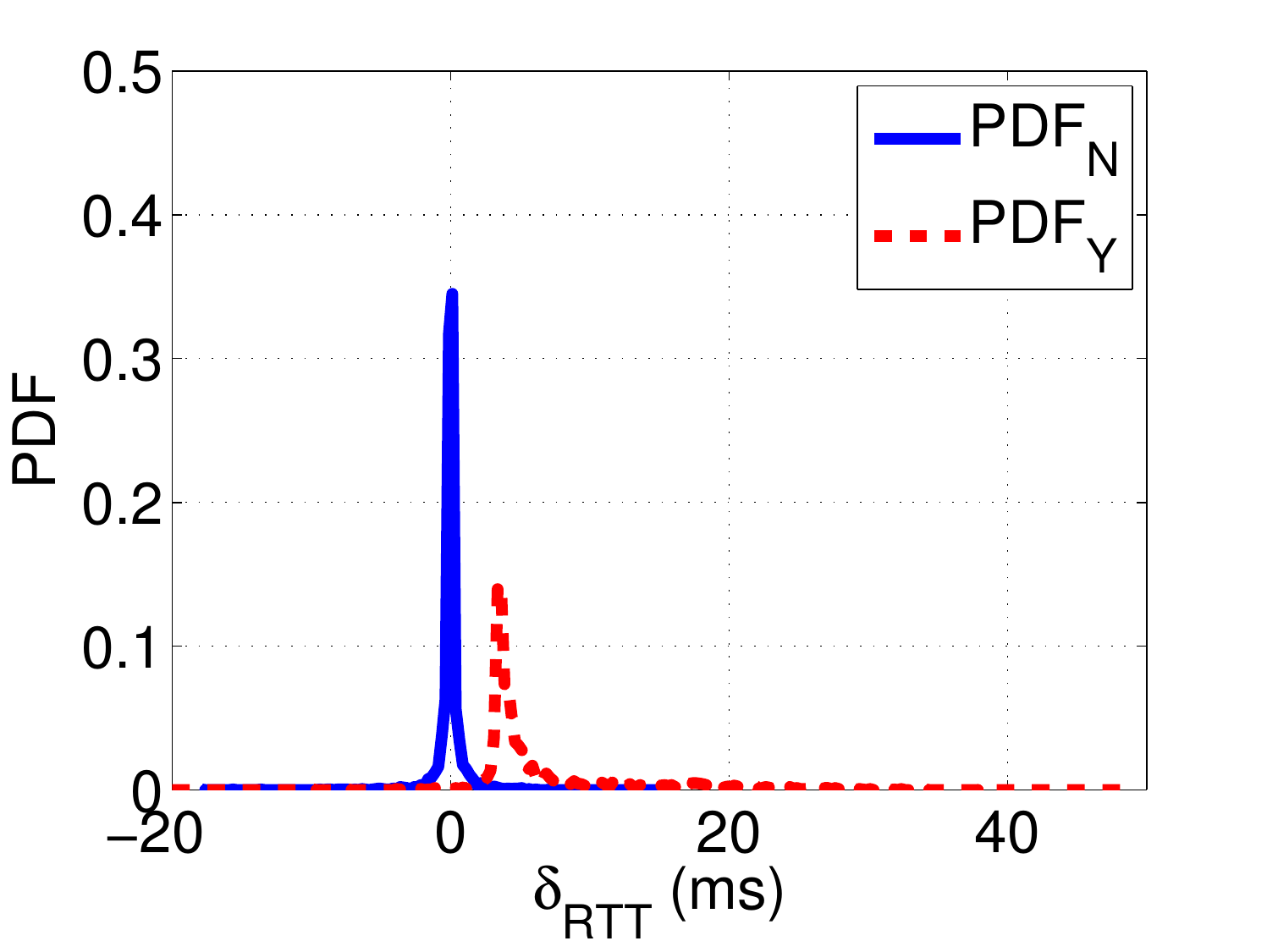}}
  \subfigure[$k=1$ software switch, time~span~1\,second]{\label{fig:delta_rtt_1soft_1gbps}\includegraphics[width=0.245\textwidth]{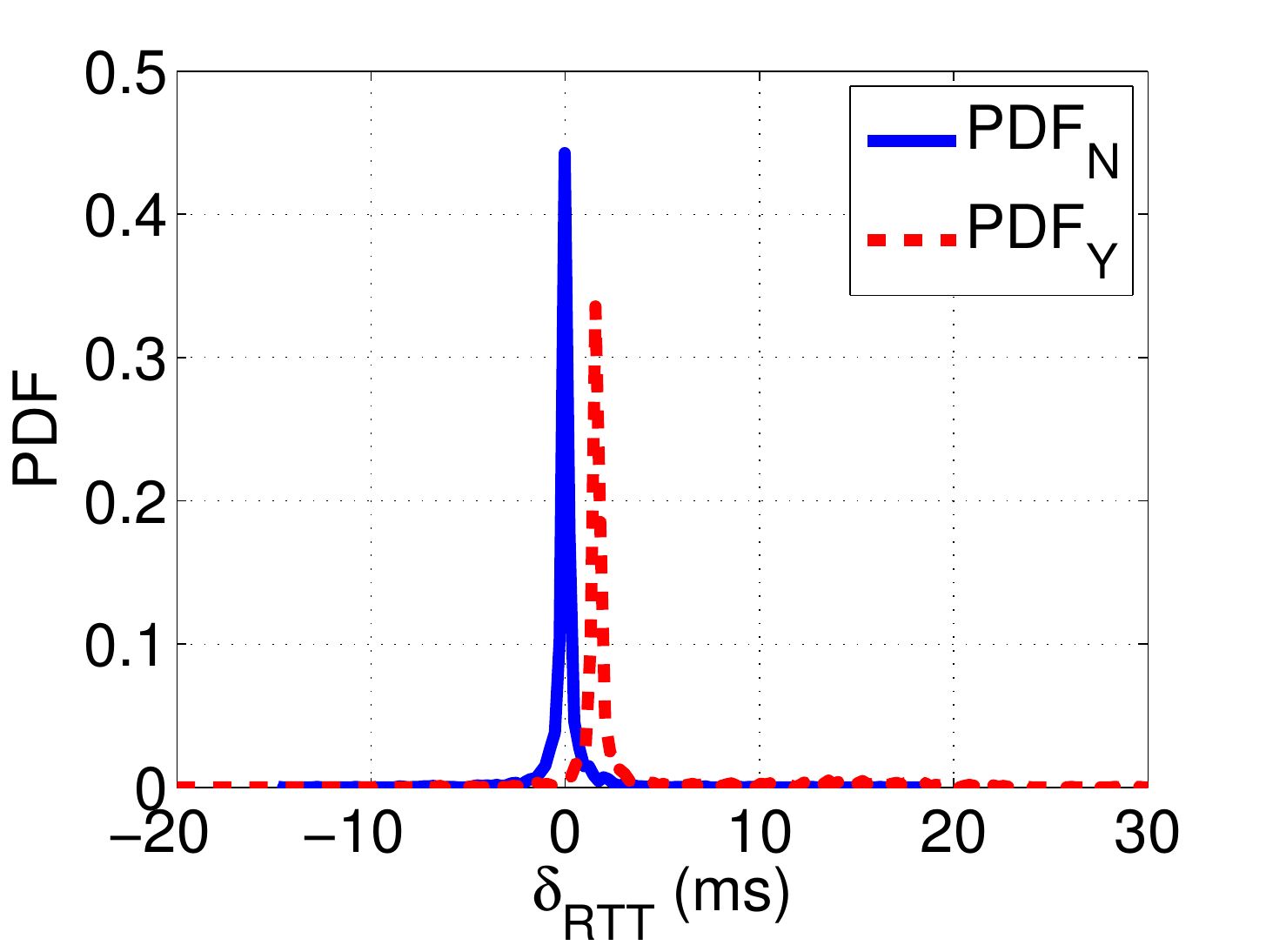}}
  \caption{Fingerprinting SDN networks (1\,Gbps link) using
    $\delta_{\mathit{RTT}}$. In our plots, we assume a bin size of
    \binSize.}
  \label{fig:pdf_rtt_1gbps}
\end{figure*}
In Figure~\ref{fig:pdf_rtt}, we plot the PDF of
$\delta_{\mathit{RTT}}$ values witnessed by probe packets sent within
a short time interval (i.e., by the last two probes of our probe train
sent 1 second apart) with respect to a varying number of switches. Our results show
that, irrespective of the number of switches, the PDF of all $\delta_{\mathit{RTT}}$ values
collected by our clients for which neither of the two probes triggered
any rule installation on the switches (referred to as
$\mathit{PDF}_N$) can be fitted to a normal distribution with mean
0. In contrast, the sample mean of PDF of $\delta_{\mathit{RTT}}$
values for which only the first probe ($\mathit{PDF}_Y$) incurred a
rule installation is strictly greater than 0; $\mathit{PDF}_Y$ and
$\mathit{PDF}_N$ are significantly different at 1\% according to
t-test. The EER is approximately 0.43\% when $k=3$ hardware switches, 0.13\% when $k=2$ hardware switches,
1.25\% when $k=1$ hardware switch, and increases to 5.84\% when $k=1$ software switch.

Similar to the dispersion feature, our results are little affected by
the speed of the data link (cf. Figure~\ref{fig:pdf_rtt_1gbps}). More specifically, the EER was unchanged
for $k=3$ hardware switches and marginally increased by almost 0.7\%
for $k=2,1$ hardware switches when the bandwidth of the data link
increases from 100\,Mbps to 1\,Gbps. Given this change, the EER
increased by almost 1.5\% when $k=1$ software switch.

\begin{figure}[t]
  \centering
  \includegraphics[width=0.317 \textwidth]{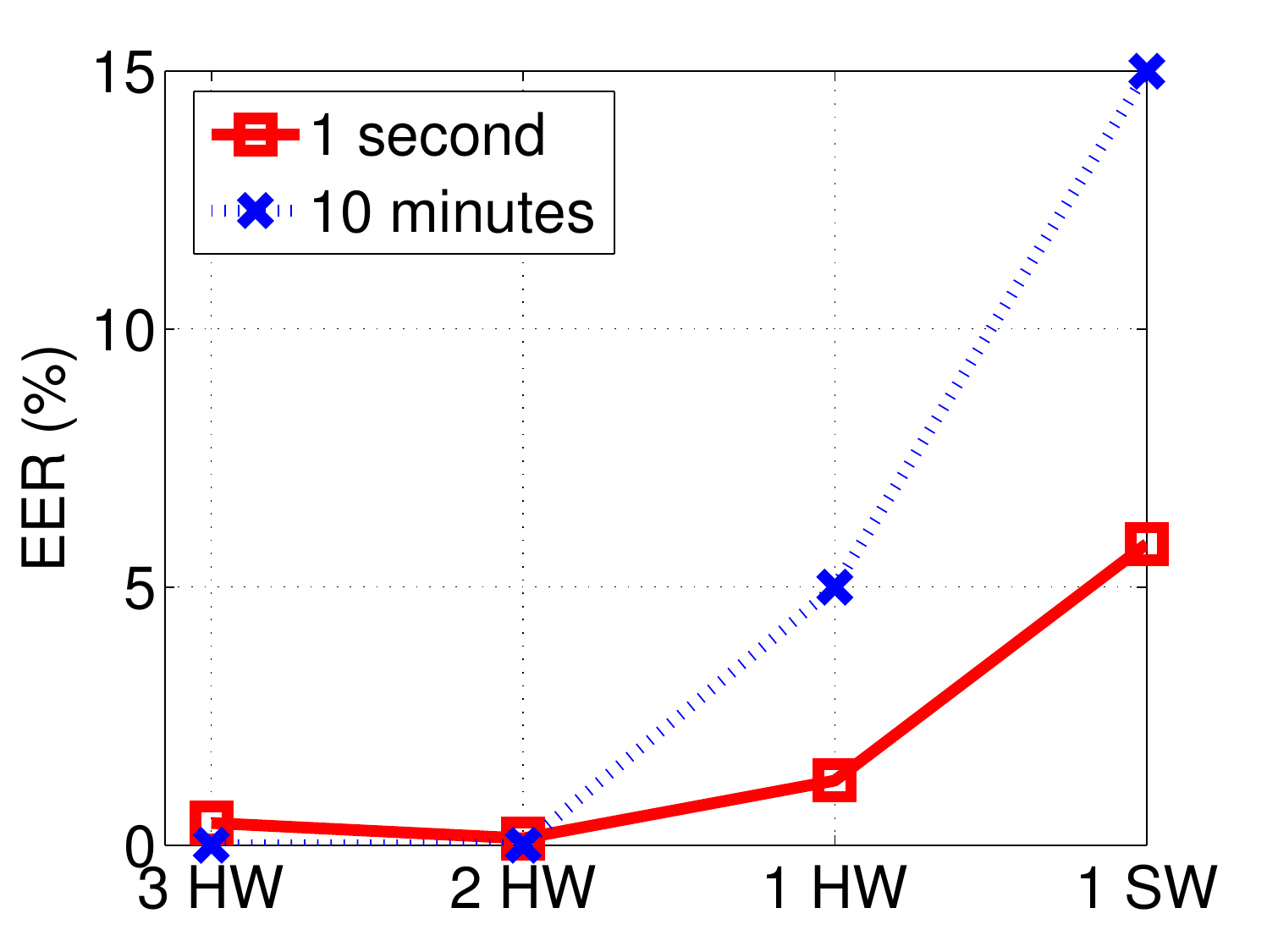}
  \caption{Impact of the time span on the $\delta_{\mathit{RTT}}$ feature.}
  \label{fig:eer_different_period}
\end{figure}

\begin{figure}[t]
\centering
\includegraphics[width=0.32 \textwidth]{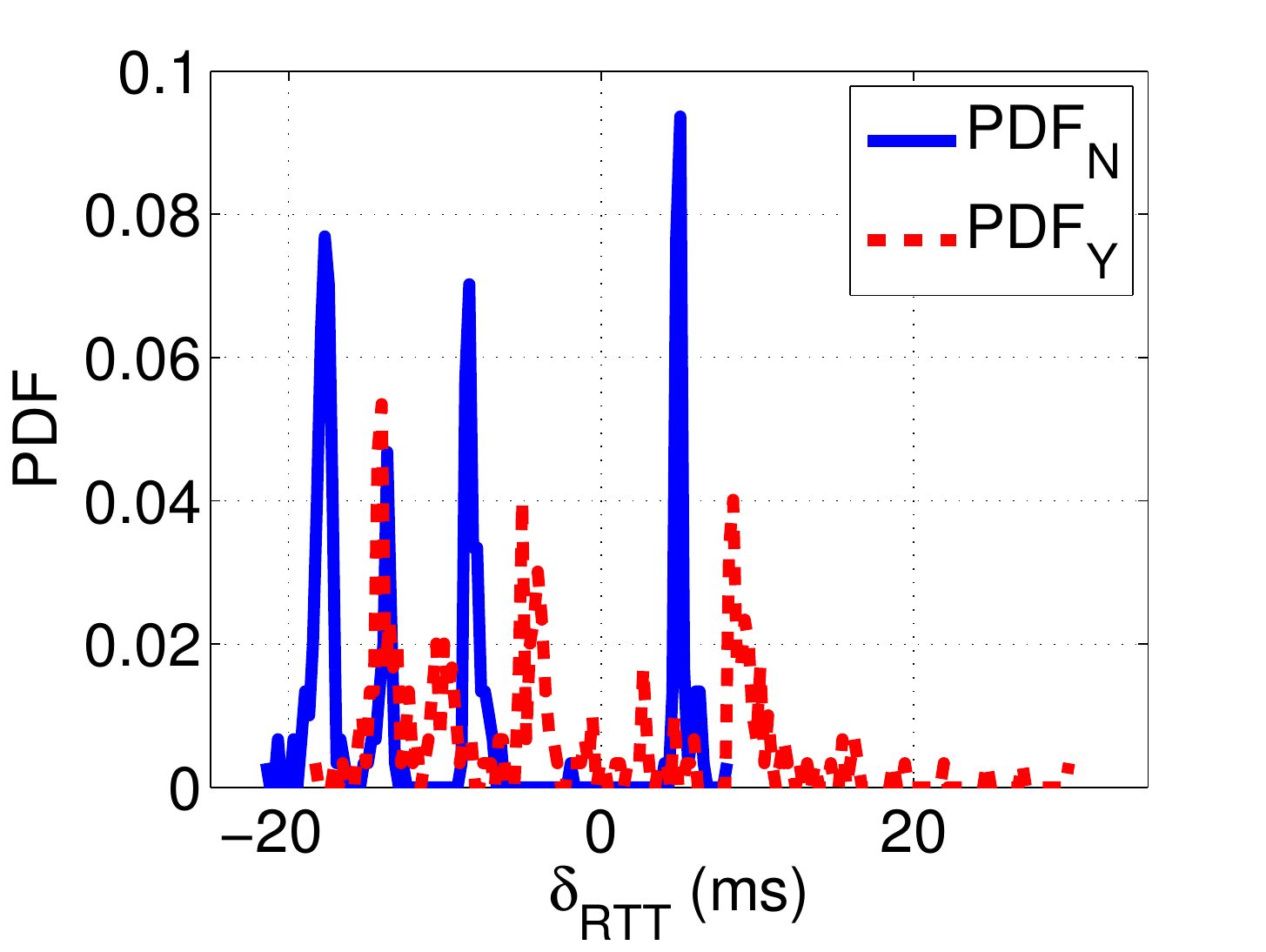}
\caption{$\delta_{\mathit{RTT}}$ over a period of 3 months when $k=1$ hardware switch. Here, we assume a bin size of \binSize.}
\label{fig:delta_rtt_10}
\end{figure}

We also compute the EER for $\delta_{\mathit{RTT}}$ values measured
over a 10 minute span. As shown in Figure~\ref{fig:eer_different_period}, our
results indicate that $\delta_{\mathit{RTT}}$ is not a stable feature
over time; for example, when $k=1$, the EER deteriorates
to approximately 5\% for a hardware switch, and almost 15\% when dealing with a software switch.
To further study the stability of $\delta_{\mathit{RTT}}$ over a longer period of time (three months), we conducted a separate experiment using four of our nodes located in Europe. Our results (cf.
Figure~\ref{fig:delta_rtt_10}) show that
$\mathit{PDF}_N$ and $\mathit{PDF}_Y$ are considerably less distinguishable when RTT
values are collected over a larger time span; for example, the EER
grows to 39.5\% for a time span of 3 months when $k=1$ hardware switch (and to 25.17\% and 24.83\% when $k=2$ and $k=3$, respectively). We believe that this discrepancy is due to changing network conditions, which incur in non-negligible differences in RTT~\cite{Schwartz:2010:MSO:1889324.1889327}. Namely, our results suggest that the change of the RTT value by a few milliseconds, caused by e.g., a change in the WAN path or traffic conditions, is comparable to the change in the RTT value introduced by an interaction with the controller.

\subsection{Summary of Results}\label{subsec:impl}

Our evaluation results in Figures~\ref{fig:pdf_disp},~\ref{fig:pdf_disp_1gbps},~\ref{fig:pdf_rtt}, and~\ref{fig:pdf_rtt_1gbps} show that fingerprinting attacks on SDN networks
are feasible; in fact, they are already realizable using simple features such as packet-pair dispersions and RTTs.

More specifically, our findings suggest that, irrespective of the number of OpenFlow switches that need to be configured in reaction to a given probe packet, the delay introduced by rule installation, $\max_{\forall k}\delta^i_{k}$, provides an effective distinguisher for an adversary to identify whether packets are only processed on the fast data plane, or triggers an interaction with the controller on the relatively slow software-based control plane. This delay is clearly distinguishable using the packet-pair dispersion, which is a stable feature over time, and is little affected by the size of the network (i.e., by the number of OpenFlow switches that need to be configured).

Although packet pairs can be easily crafted by an \emph{active} adversary, packet pairs might not always be extractable from existing traffic by a \emph{passive} adversary.
However, a passive adversary can monitor existing traffic for packets that share a similar packet header, and are sent apart within a short time interval (e.g., within 10 minutes).

Our findings show that the difference of measured RTT values between two such packets provides evidence for a passive adversary whether any of those packets triggered a reaction on the control plane. Passive fingerprinting is especially detrimental since it limits the applicability of existing intrusion detection systems in detecting SDN fingerprinting attempts; indeed, passive network monitoring does not generate any extra traffic and as such cannot be deterred by relying on anomaly detection.
Notice, however, that the RTT feature considerably depends on the SDN
network size and is less stable over time when compared to packet-pair
dispersions. For instance, the EER almost
doubles in our experiments when the number of OpenFlow hardware switches that need to be configured decreases from $3$ to $1$.

Our results also suggest that the data link bandwidth has little impact on the fingerprinting accuracy for both the RTT and the dispersion features. The presence of a software switch in the communication path, however, considerably deteriorates the EER; even in such a setting, our results nevertheless show that fingerprinting attacks can still be reliably mounted by a remote adversary.

\begin{figure}[t]
  \centering
  \subfigure[Dispersion feature.]{\label{fig:bar_dispersion}\includegraphics[width=0.317 \textwidth]{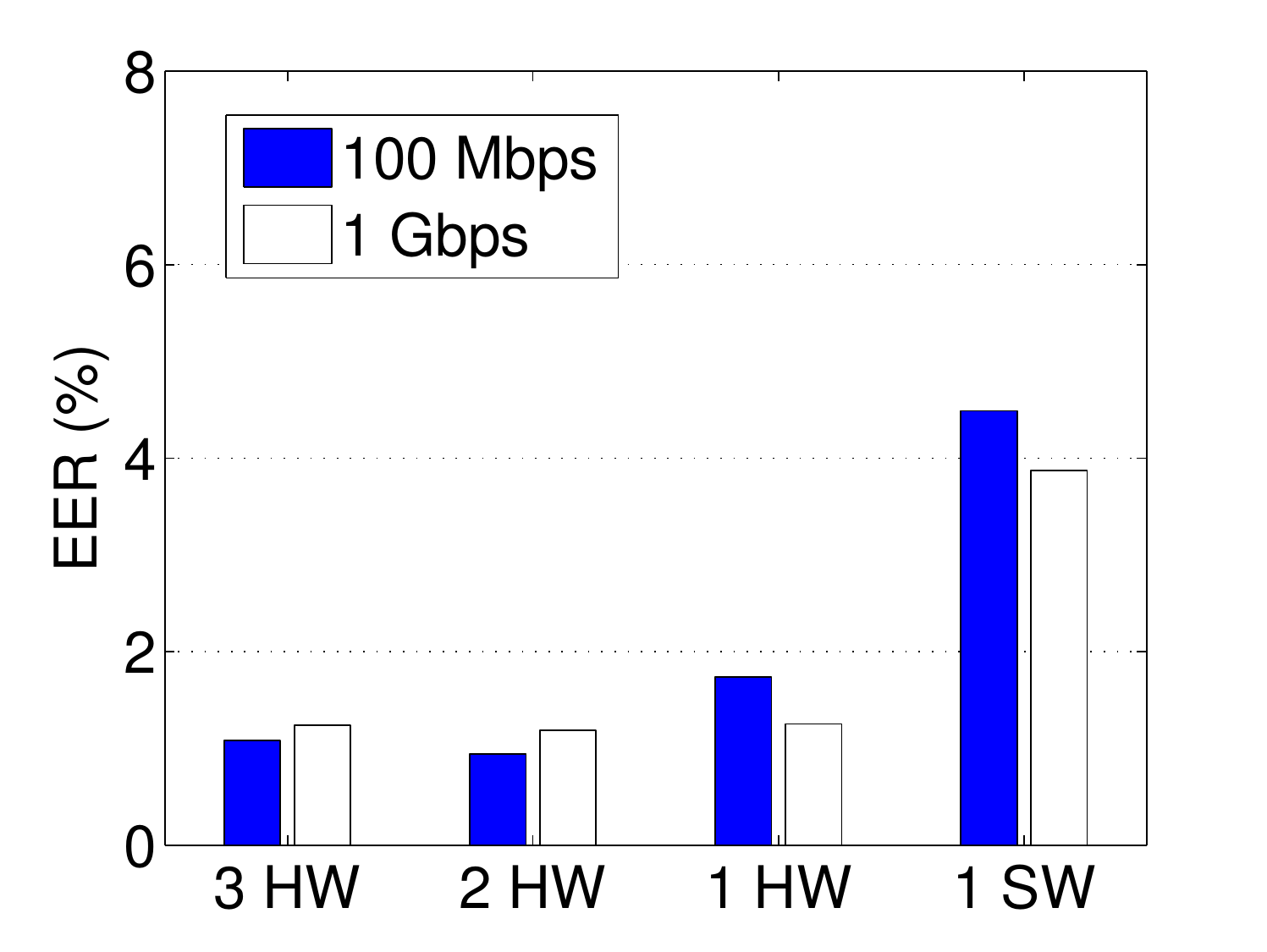}}
  \subfigure[$\delta_{\mathit{RTT}}$ feature.]{\label{fig:bar_deltaRTT}\includegraphics[width=0.317 \textwidth]{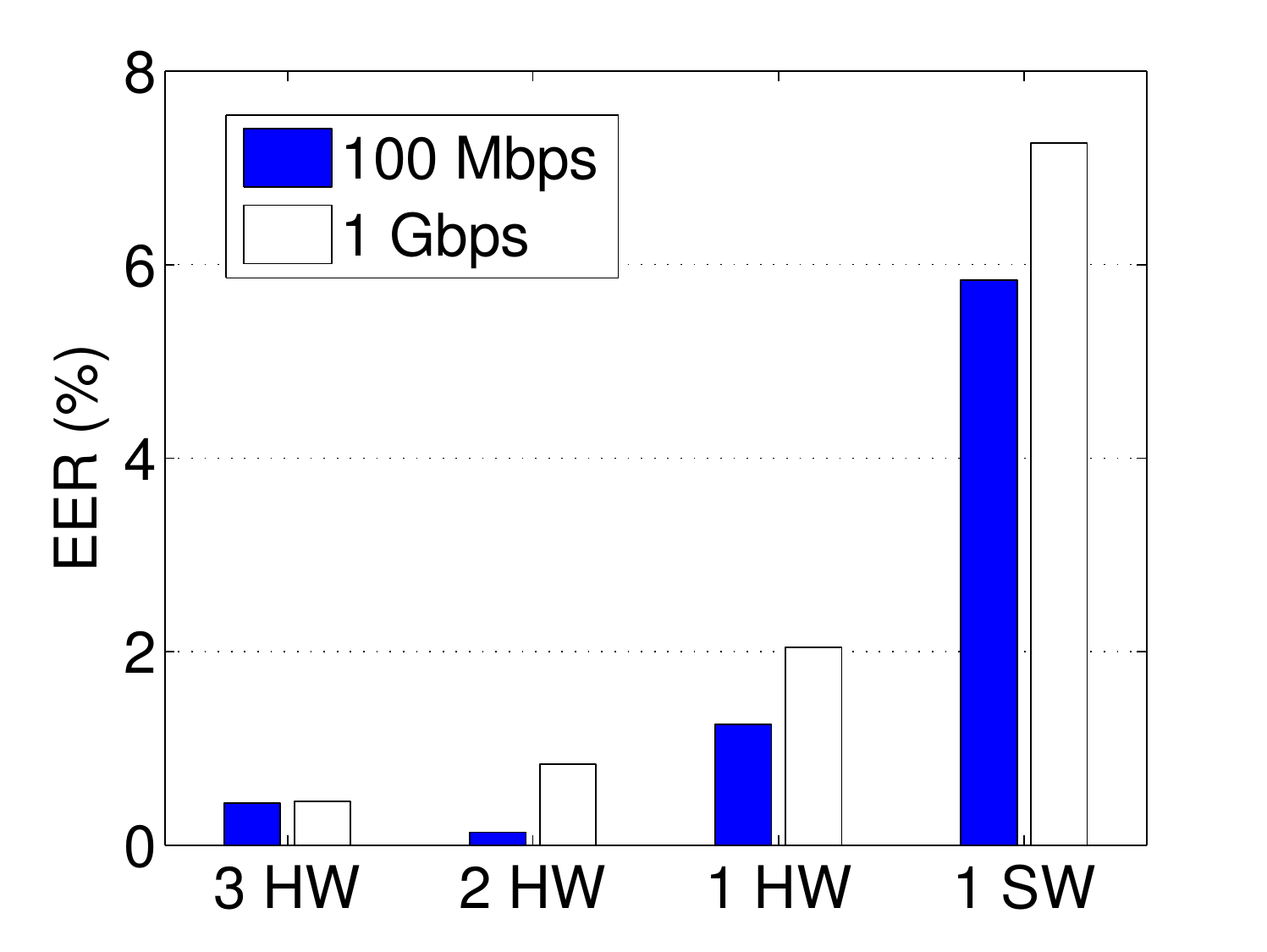}}
  \caption{Fingerprinting SDN networks with respect to the number of switches and the data link bandwidth.}
  \label{fig:bar}
\end{figure}

\begin{table*}[t]
\centering
\caption{Summary of obtained EERs.}
\label{tab:summary}
\scalebox{0.9}{
\begin{tabular}{|c|c|c|cccc|cccc|}
\hline
\multicolumn{3}{ |c|  }{}
& \multicolumn{4}{ |c|  }{100\,Mbps data link} & \multicolumn{4}{ |c|  }{1\,Gbps data link}\\
\cline{4-11}
\multicolumn{3}{ |c|  }{}&
$k=3$ HW & $k=2$ HW & $k=1$ HW & $k=1$ SW &
$k=3$ HW & $k=2$ HW & $k=1$ HW & $k=1$ SW\\
\hline\hline
\multicolumn{2}{|c|}{\multirow{2}{*}{Packet-pair Dispersion}}
&
  EER       &  1.08\%          & 0.94\%          & 1.74\%  & 4.49\% & 1.24\%          & 1.19\%          & 1.25\%  & 3.87\% \\
\multicolumn{1}{|c}{} & &
  Threshold & 1.43\,ms & 1.37\,ms & 1.17\,ms  &  0.37\,ms & 1.42\,ms & 1.36\,ms & 0.88\,ms  &  0.20\,ms
      \\
\hline\hline
\multicolumn{2}{|c|}{\multirow{2}{*}{$\delta_{\mathit{RTT}}$,
    time span 1 second}} &
 EER       &  0.43\%          & 0.13\%          & 1.25\%  & 5.84\% &  0.45\%          & 0.84\%          & 2.04\%  & 7.25\% \\
\multicolumn{1}{|c}{}
& &
 Threshold & 4.63\,ms & 4.27\,ms & 2.13\,ms  &  0.84\,ms & 4.60\,ms & 4.85\,ms & 2.18\,ms  &  0.85\,ms
 \\
\hline
\end{tabular}}
\end{table*}
Table~\ref{tab:summary} and Figure~\ref{fig:bar} summarize our results. We argue that our findings apply to other SDN networks in which the relative difference between the processing speed of packets at the data plane, and at the control plane is even more pronounced. Recall that our testbed was devised to
emulate a scenario that is particularly hard for fingerprinting. That is, the controller's CPU was idle most of the time during the
measurements; the controller used pre-computed rules when issuing
forwarding decision and was connected to a small number of switches
(i.e., three); at the time of writing, the deployed OpenFlow hardware switches are
among the fastest in installing new flow rules; furthermore, we ensured that the switches' flow tables were empty when performing the measurements, obtaining a flow rule installation time in the order of milliseconds~\cite{devoflow,tango}.
Hence, it is clear that the fingerprinting accuracy provided by our
features only increases when the controller is under heavy load, the data plane bandwidth is larger (e.g., 10\,Gbps), or the OpenFlow switches require longer times to update their flow tables.

Note, however, that our findings rely on the assumption that there is a single SDN network on the path to the server. Otherwise, while our features
still fingerprint controller-switch interactions in more than one SDN network, they do not reveal in which network the interactions take place.

\section{Implications}\label{sec:impl}

In the previous section, we showed that remote fingerprinting attacks on SDN networks are feasible and easily realizable by means of our proposed features.
In what follows, we discuss the implications of our findings on the security of SDN networks. %

\subsection{Rule Scanning}
Based on our findings, a remote adversary can clearly infer whether a
flow rule has been already installed by the controller in order to
handle a specific type of traffic or route towards a given
destination.  For example, the adversary can craft probe packets whose
headers match the traffic type and/or destination address and infer by
measuring the timing of the packets whether these packets triggered
the installation of a rule.  This provides a strong evidence for the
adversary that e.g., communication with the given destination address
has recently occurred. Depending on the underlying rule, the adversary
might also be able to infer the used network protocol, and the
destination port address. By doing so, the adversary obtains
additional information about the occurrence of a particular
communication event; for example, the adversary can infer whether the
destination address has recently established an SSL session to perform
an e-banking transaction. Notice that this leakage is only particular
to SDN networks, and does not apply to traditional networks.

Moreover, the remote fingerprinting of rules enables the adversary to
better understand the logic adopted by the controller in managing the
SDN network. This includes inferring the timeouts set for the expiry
of specific rules, whether the controller aims at fine-grained or
coarse-grained control in the network, etc. Similar to existing port
and traffic scanners, this knowledge can empower the adversary with
the necessary means to compromise the SDN network. Even
worse, the adversary can leverage this knowledge in
order to attack other networks which implement a similar rule installation logic. For instance, in a geographically dispersed datacenter, different sub-domains typically implement the same policies. The adversary can train using one sub-domain and leverage the acquired knowledge in order to compromise another sub-domain.

\subsection{Denial-of-Service Attacks}
The rule space is a scarce resource in existing hardware switches. Namely,
state-of-the-art OpenFlow hardware switches can only accommodate few tens of
thousands rules~\cite{cacheflow}, and only support a limited number of
flow-table updates per second~\cite{tango, Bifulco2014}.
While these limitations can be circumvented by means of a careful design
of the rule installation logic, an adversary that knows which packets
cause an interaction with the controller can abuse this knowledge to
launch Denial-of-Service (DoS) attacks.

For instance, an adversary might simply try to overload the controller
with handling \textit{packet-in} events.  More specifically, the
adversary sends packets, where each of them most likely triggers a
controller-switch interaction.  Too many such interactions will
overload the controller.

Another kind of DoS attack is to fill up the switches' flow tables.
An
analogy to this is when a computer runs out of memory and starts
swapping. Usually, the computer becomes unusable. Similarly, the network performance is severely harmed when the flow tables are full (or even almost full).
First, installing flow rules in an almost full table is more costly
than in an almost empty flow table. Second, in case the flow table is
full, either new network flows cannot be established, which would
already be a DoS, or some installed flow rules need to be deleted.
However, in general, it is not obvious which rules should be deleted
to make room for new rules; this needs to be coordinated
by the controller and is a complex operation, which can quickly
overload the controller and the switches.  For example, the deletion
of a rule of an ongoing network flow might entail the rule's
immediate reinstallation. This can escalate and the controller will have to
constantly delete and reinstall rules.

An adversary can make both kinds of DoS attacks more likely to succeed
by first passively fingerprinting the network traffic, instead of
blindly guessing which packets trigger a controller-switch
interaction.

\section{Proposed Countermeasure}\label{sec:countermeasure}

In this section, we present an efficient countermeasure to prevent fingerprinting attacks on SDN networks. We also evaluate the effectiveness of our countermeasure in our testbed.

\subsection{Our Countermeasure}\label{subsec:countermeasure}

One possible countermeasure against fingerprinting would be for the swit\-ches
to delay every received packet before forwarding it.  This
countermeasure is clearly inefficient as it would severely harm
network performance.  Randomly deleting flow rules and reinstalling
them when receiving the corresponding \textit{packet-in} events also
does not solve the problem either. First, additional interactions
between the controller and the switches are introduced, resulting in
an additional burden on the controller.  Second, installing flow rules
is a costly switch operation.  In what follows, we sketch an efficient countermeasure, which relies on only delaying the first few
packets of a network flow.

Our proposal does not concern the handling of new flows, but focuses on processing packets which pertain to existing flows.
Namely, for a packet of an existing flow, we
leverage the group table~\cite{OpenFlow_v1.3.0} and the
internal timer maintained by a switch to identify whether
this flow has recently
appeared.
The group tables are used in OpenFlow switches to describe per-packet forwarding actions. They allow one to realize forwarding strategies, such as ECMP~\cite{ecmp}, which could not be achieved using the flow table abstraction that describes only per-flow forwarding actions. A group table contains one or more \textit{buckets}, which in turn contain an action set, similar to the one contained in the flow rules. A group table is further associated with a bucket selection logic, which is related to the group table \textit{type}. For example, a group table of type ``fast failover'' implements a selection logic that associates each bucket to a switch's port. Then, the logic selects the first bucket in the table whose associated switch's port status is \textit{live}.
The action \textit{group} in a flow rule's action set enables one at selecting which packets should be processed by which group.

\begin{figure*}[t]
  \centering
  \includegraphics[width=0.57\linewidth]{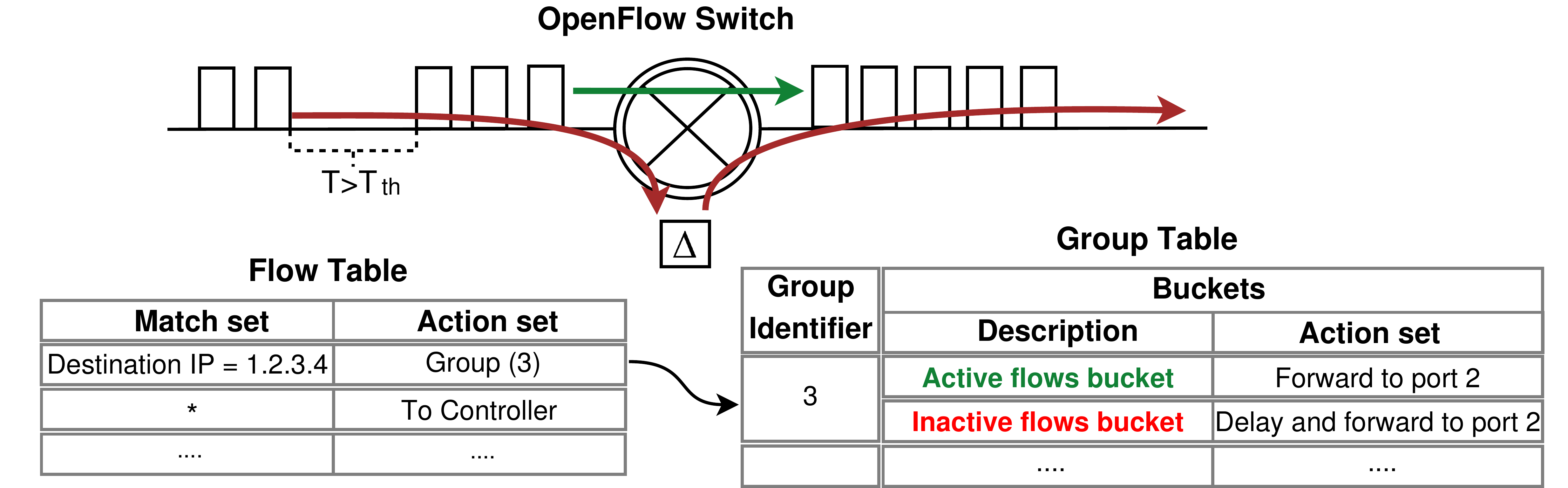}
  \caption{Sketch of our countermeasure. The packets with destination IP 1.2.3.4 are processed by the group table 3, which implements the bucket selection logic specified by our countermeasure. If no packets for this network flow are processed by the switch for a time $T > T_{th}$, then the next few packets of the flows are delayed by an appropriate amount.}
  \label{fig:countermeasure}
\end{figure*}
Our proposed countermeasure (cf. Figure \ref{fig:countermeasure}) defines a new bucket selection logic for the group table, such that packets of active flows are immediately forwarded, while packets
of inactive flows are forwarded onto a special port that connects the switch to a network delay element. Our selection logic considers a flow to be inactive if no packets for such a flow were received by the switch in a threshold amount of time, $T_{th}$, which is measured (in seconds) by the switch's internal timer.

The first received packet of an inactive flow is delayed by $\delta_{\mathit{RTT}} \approx  \max_{\forall k}\delta^i_{k}$, which gives
the adversary little
advantage in identifying whether the additional delay measured by the RTT feature is caused
by a controller-switch interaction or is artificially introduced by our countermeasure. Moreover, all packets of the same flow received within a short time window $W$ are also delayed by a small $\Delta$; this
procedure prevents fingerprinting attempts that leverage the dispersion feature. As shown in Section~\ref{subsection:evalutationcountermeasure}, $\Delta$ and $\delta_{\mathit{RTT}}$ can be fitted to pre-determined distributions, depending on the network size and the number of hops on the communication path. Alternatively, the controller can estimate the distributions corresponding to $\Delta$ and $\delta_{\mathit{RTT}}$ through a feedback loop.

Notice that our countermeasure is unlikely to deteriorate network
performance, since
only few packets per flow are delayed by few milliseconds (cf. Section~\ref{subsection:evalutationcountermeasure}). We further remark that our proposal requires minor modifications---which are supported to a large extent
 already in the OpenFlow v1.3 specification---by the switches' manufacturers. As such, we argue that our proposal can be efficiently implemented (in hardware) within the switches.

\subsection{Evaluation}\label{subsection:evalutationcountermeasure}

\begin{figure*}[t]
  \centering
  \includegraphics[scale=0.08]{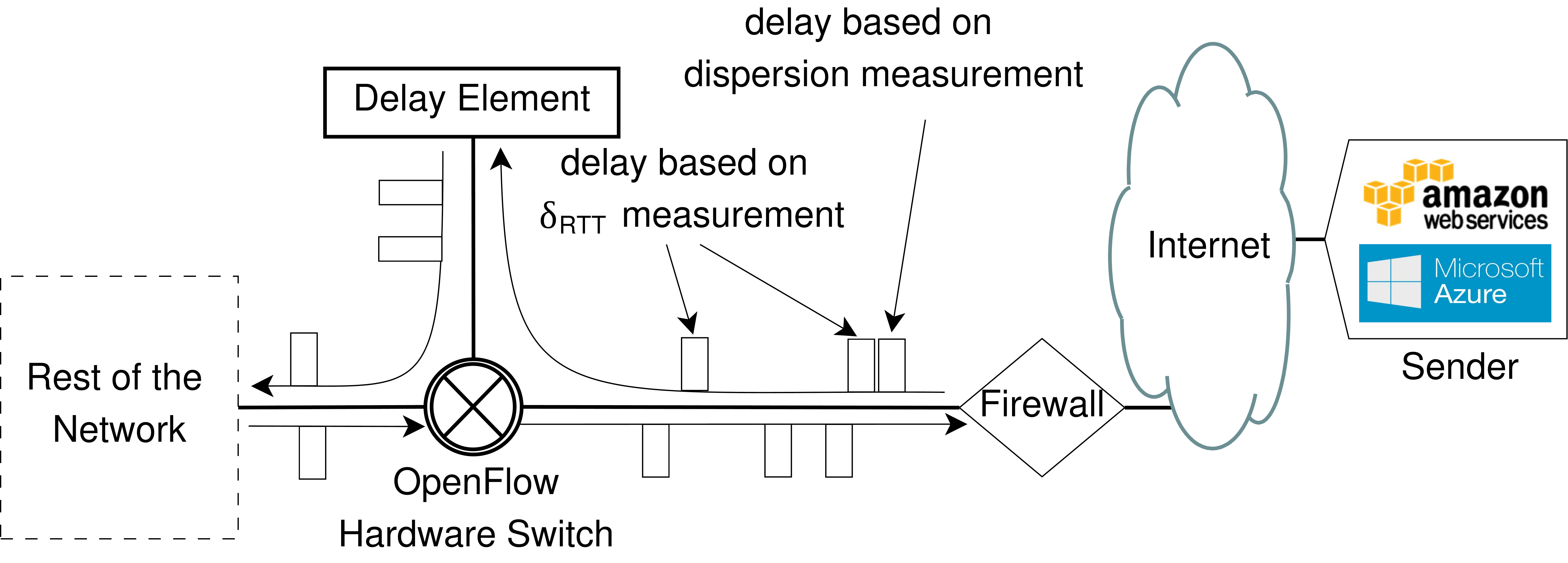}
  \caption{Modified testbed used to evaluate our countermeasure.}
  \label{fig:countermeasure_topo}
\end{figure*}
We evaluate the effectiveness of our proposed countermeasure using the
tested described in Section~\ref{subsec:data}. For that purpose, we
connect a delay element, running on Intel Xeon 2.8\,GHz CPU with 4\,GB
of RAM, to a reserved port on the outermost switch; as described
later, this element delays network packets by a specified amount
before outputting them back on another port of the switch
(cf. Figure~\ref{fig:countermeasure_topo}).

Since our countermeasure requires a modification to the bucket selection logic by the switches' manufacturers, we emulate this logic by tagging the first few packets of inactive flows, and by pre-installing a rule which forwards such tagged packets to the delay element.

\begin{figure}[t]
  \centering
  \subfigure[Dispersion feature.]{\label{fig:datafit1}\includegraphics[width=0.317 \textwidth]{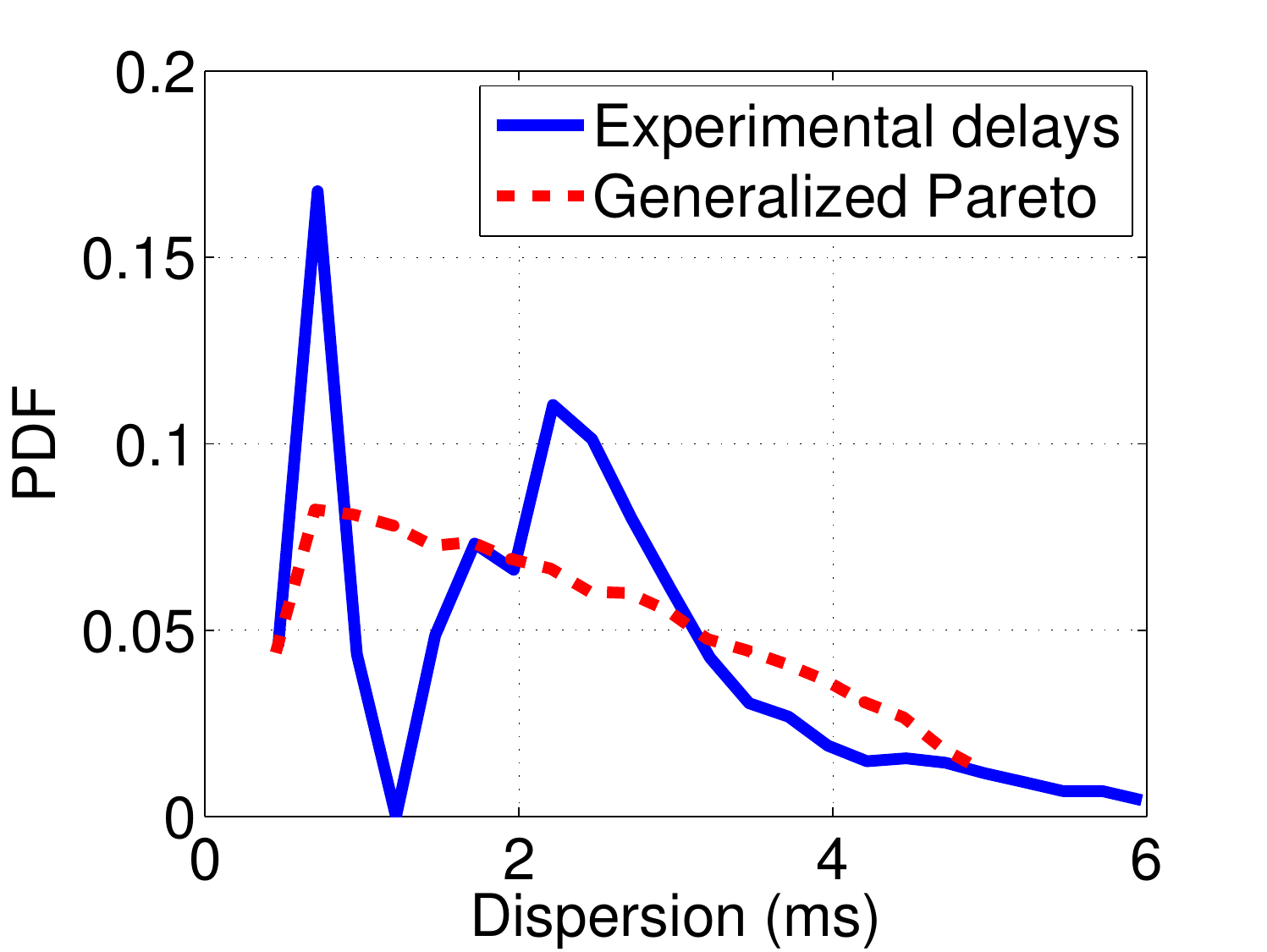}}
	  \subfigure[$\delta_{\mathit{RTT}}$ feature.]{\label{fig:datafit2}\includegraphics[width=0.317 \textwidth]{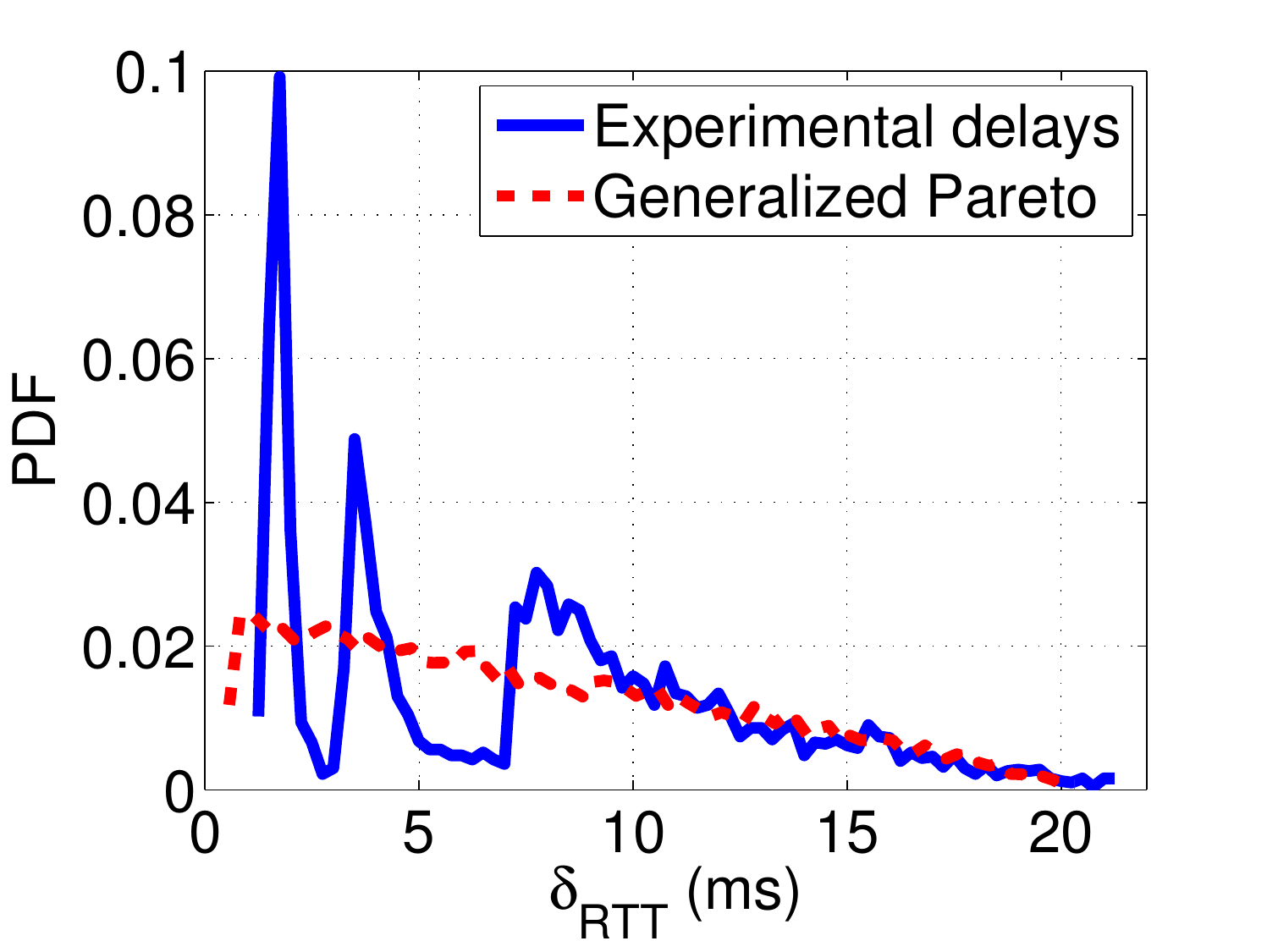}}
  \caption{Fitting experimental data to the Generalized Pareto distribution.}
  \label{fig:datafit}
\end{figure}
Packets are delayed by the delay element according to a pre-determined distribution. To select the best-fit distribution, we applied the Kolmogorov-Smirnov test~\cite{ksTest} on a number of candidate distributions, such as Pareto, Generalized Pareto, Weibull, using our collected measurements as ground truth. Our results show that the Generalized Pareto distribution achieves the highest Kolmogorov-Smirnov score test (of approximately 0.09 for dispersion and 0.08 for $\delta_{\mathit{RTT}}$) for both our investigated features (cf. Figure~\ref{fig:datafit}). Recall that the Generalized Pareto distribution is of the form:
\begin{equation*}
f_{(\xi,\mu,\sigma)}{(x)} = \frac{1}{\sigma}\big(1 + \frac{\xi(x-\mu)}{\sigma}\big)^{(-\frac{1}{\xi}-1)}
\end{equation*}

Table~\ref{tab:countermeasure_genPareto_parameters} summarizes the parameters for Generalized Pareto distributions extracted from our measurements, and used by the delay element to prevent fingerprinting attempts using the dispersion and the $\delta_{\mathit{RTT}}$ features. Namely, in our countermeasure, the delay element inserts a delay before transmitting the first packet by randomly sampling from a Generalized Pareto distribution with parameters $\xi=-0.53, \sigma=10.58$, and $\mu=0.57$, while all subsequent packets sent within an interval of $100$~ms are delayed by randomly sampling from a Generalized Pareto distribution with parameters
$\xi=-0.60, \sigma=2.86$, and $\mu=0.45$ (to prevent fingerprinting using the dispersion feature).
\begin{table}[t]
  \centering
  \caption{Parameters of the Generalized Pareto distributions used by the delay element.}
  \label{tab:countermeasure_genPareto_parameters}
  \scalebox{0.9}{\begin{tabular}{|c|c|c|c|c|c|}
      \hline
        &   $ \xi $ & $ \sigma $ & $ \mu $
      \\
      \hline\hline
      {Packet-pair Dispersion}
      &  -0.60      &  2.86    &  0.45  \\
			\hline
      {$\delta_{\mathit{RTT}}$}
      &   -0.53     &  10.58    &  0.57  \\
      \hline
    \end{tabular}}
\end{table}

\paragraph*{\normalfont\textbf{Results}}

\begin{table}[tb]
  \centering
  \caption{Summary of measured EERs using our countermeasure.}
  \label{tab:summary_countermeasure_test1}
  \scalebox{0.9}{\begin{tabular}{|c|cccc|}
      \hline
        &  $k=3$ HW & $k=2$ HW & $k=1$ HW & $k=1$ SW
      \\
      \hline\hline
      {Packet-pair Dispersion}
         &    39.76\%      &  46.25\%      &  61.18\%  &  84.57\%  \\ \cline{1-5}
			\cline{1-5}
		 {$\delta_{\mathit{RTT}}$}
         &  33.42\%     &   37.48\%   & 67.19\% &  83.11\%   \\
      \hline
    \end{tabular}}
\end{table}
In the remainder of this section, we report on the effectiveness of our proposed countermeasure using the testbed shown in Figure~\ref{fig:countermeasure_topo}.
Here, we collect our measurements using the same process described in Section~\ref{subsec:data}. Our results are shown in Table~\ref{tab:summary_countermeasure_test1}.

\begin{figure}[tb]
  \centering
  \subfigure[Dispersion feature.]{\label{fig:datafit_onlyOVS_1}\includegraphics[width=0.317 \textwidth]{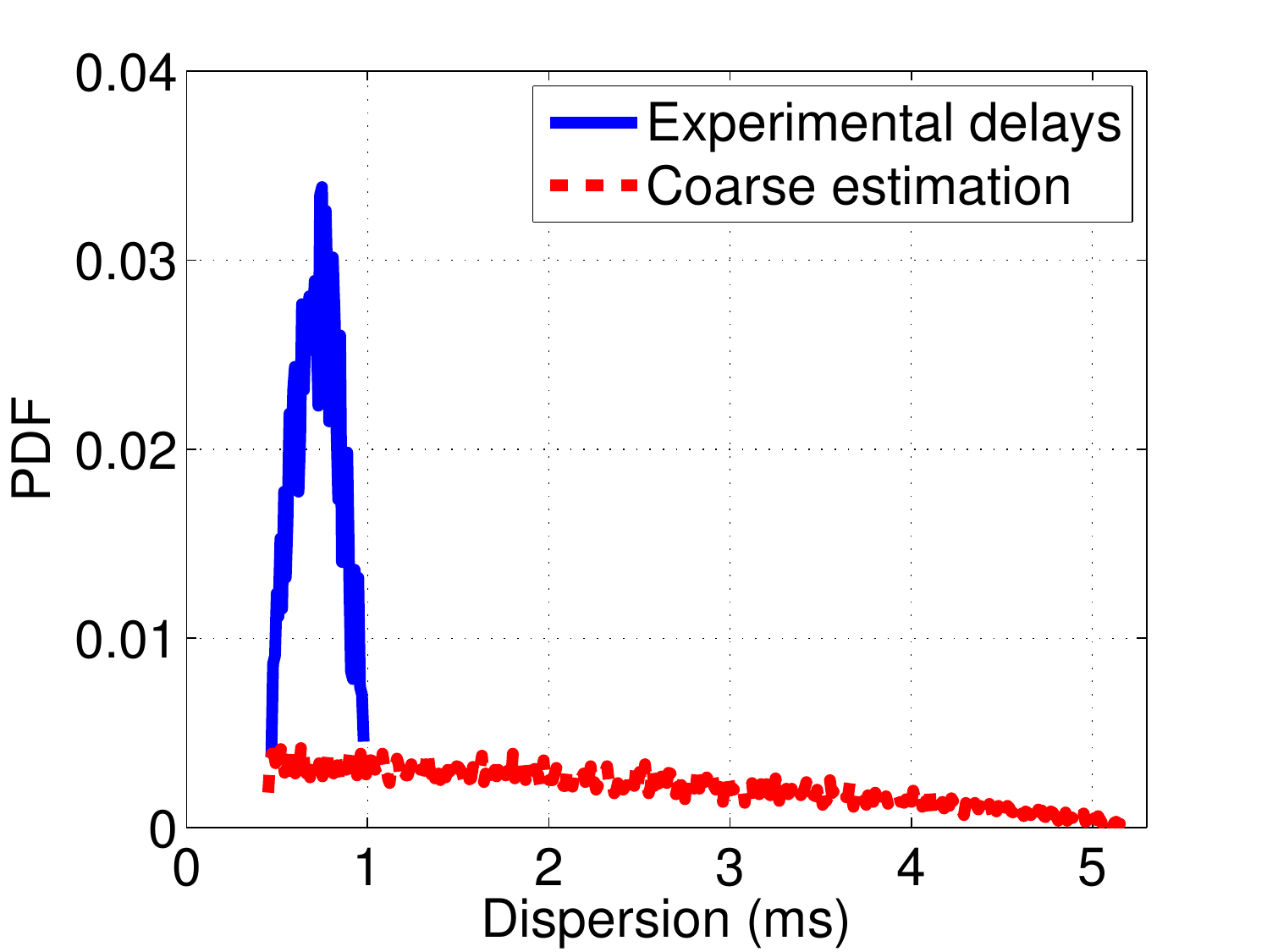}}
	  \subfigure[$\delta_{\mathit{RTT}}$ feature.]{\label{fig:datafit_onlyOVS_2}\includegraphics[width=0.317 \textwidth]{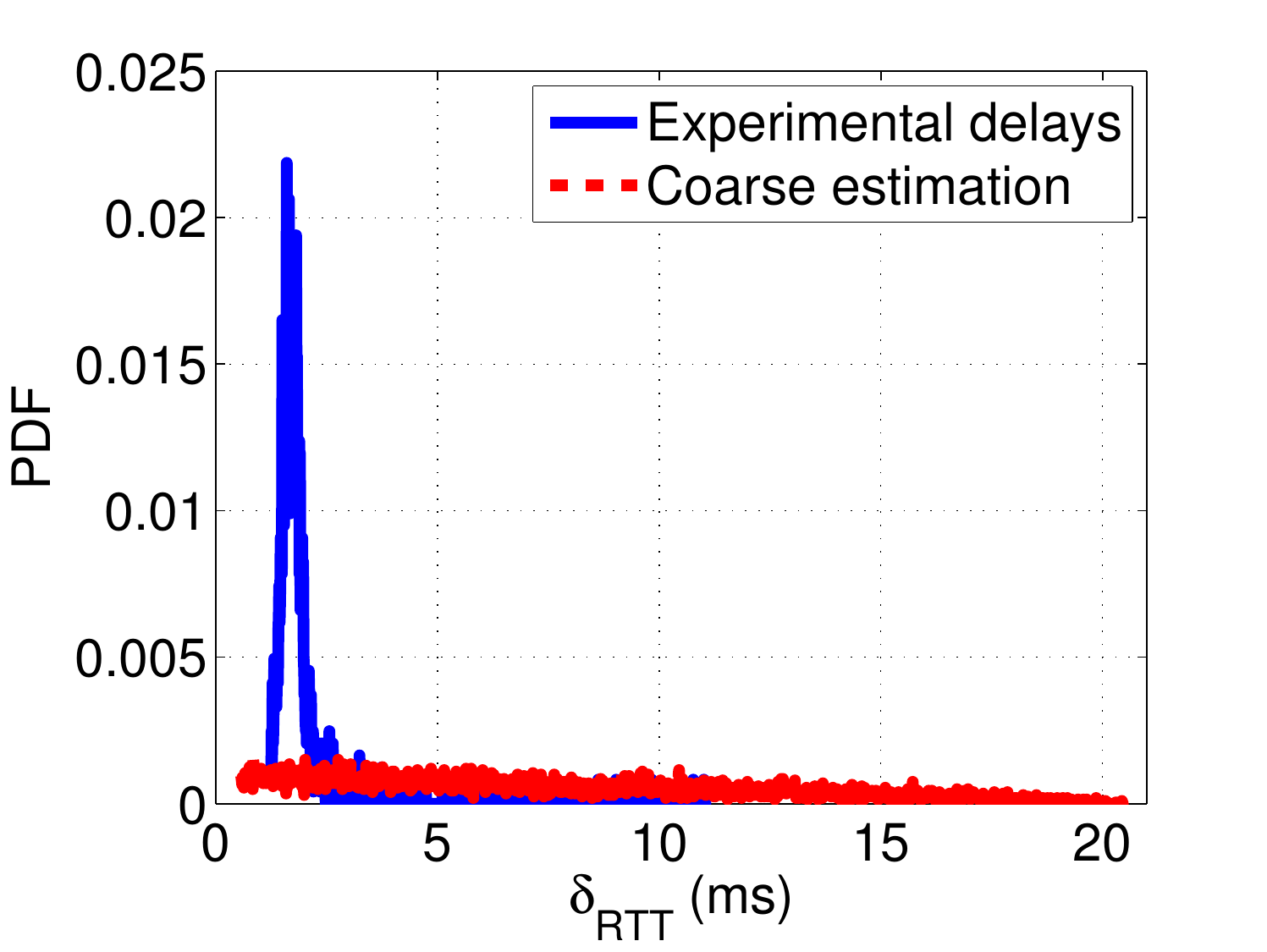}}
  \caption{Software switch experimental delays vs. the Generalized Pareto distribution.}
  \label{fig:datafit_onlyOVS}
\end{figure}

Our results indicate that our countermeasure considerably impacts the fingerprinting accuracy of a remote adversary using the dispersion and $\delta_{\mathit{RTT}}$ features. More specifically,
our countermeasure increases the EER to almost
40\% using the dispersion feature, and to 33\% using the
$\delta_{\mathit{RTT}}$ feature when the network comprises
three hardware switches. Our countermeasure, however, increases the
EER to almost 84\% (using both investigated
features) when the network comprises a software switch. Recall that the worst attainable fingerprinting
accuracy in this case is when the EER is 50\% which signals that the two distributions $PDF_Y$ and $PDF_N$ completely overlap.
In the case of a software switch, the EER increases to 84\% which means that the adversary has an advantage in distinguishing $PDF_Y$ from $PDF_N$, in spite of our countermeasure. We believe that this discrepancy mainly originates from the fact that the estimated Generalized-Pareto distribution does not emulate well delays corresponding to software switches (cf. Figure~\ref{fig:datafit_onlyOVS}).

\begin{figure}[tb]
  \centering
  \subfigure[Dispersion feature.]{\label{fig:countermeasure_dispersion}\includegraphics[width=0.317 \textwidth]{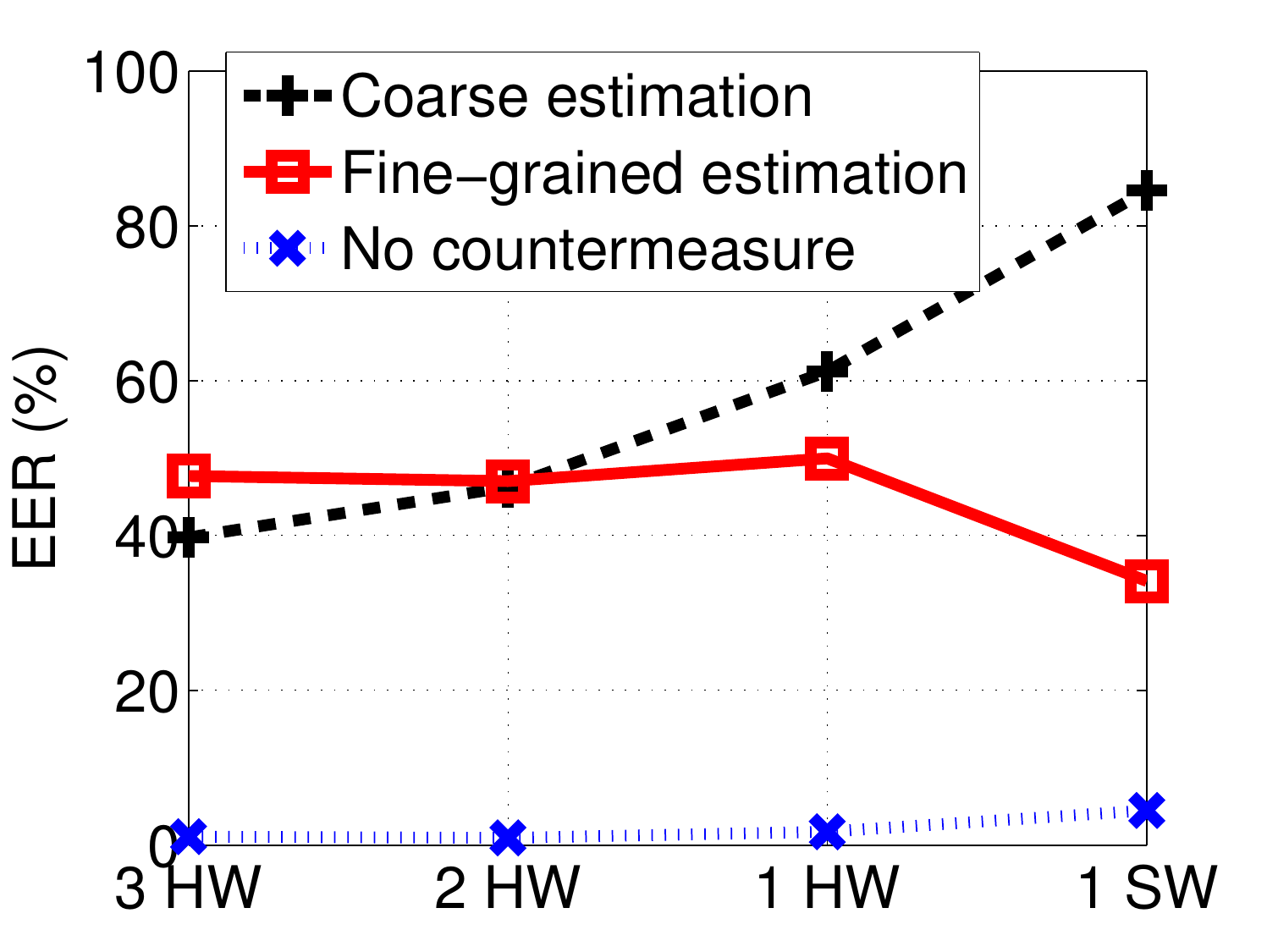}}
\subfigure[$\delta_{\mathit{RTT}}$ feature.]{\label{fig:countermeasure_deltaRTT}\includegraphics[width=0.317 \textwidth]{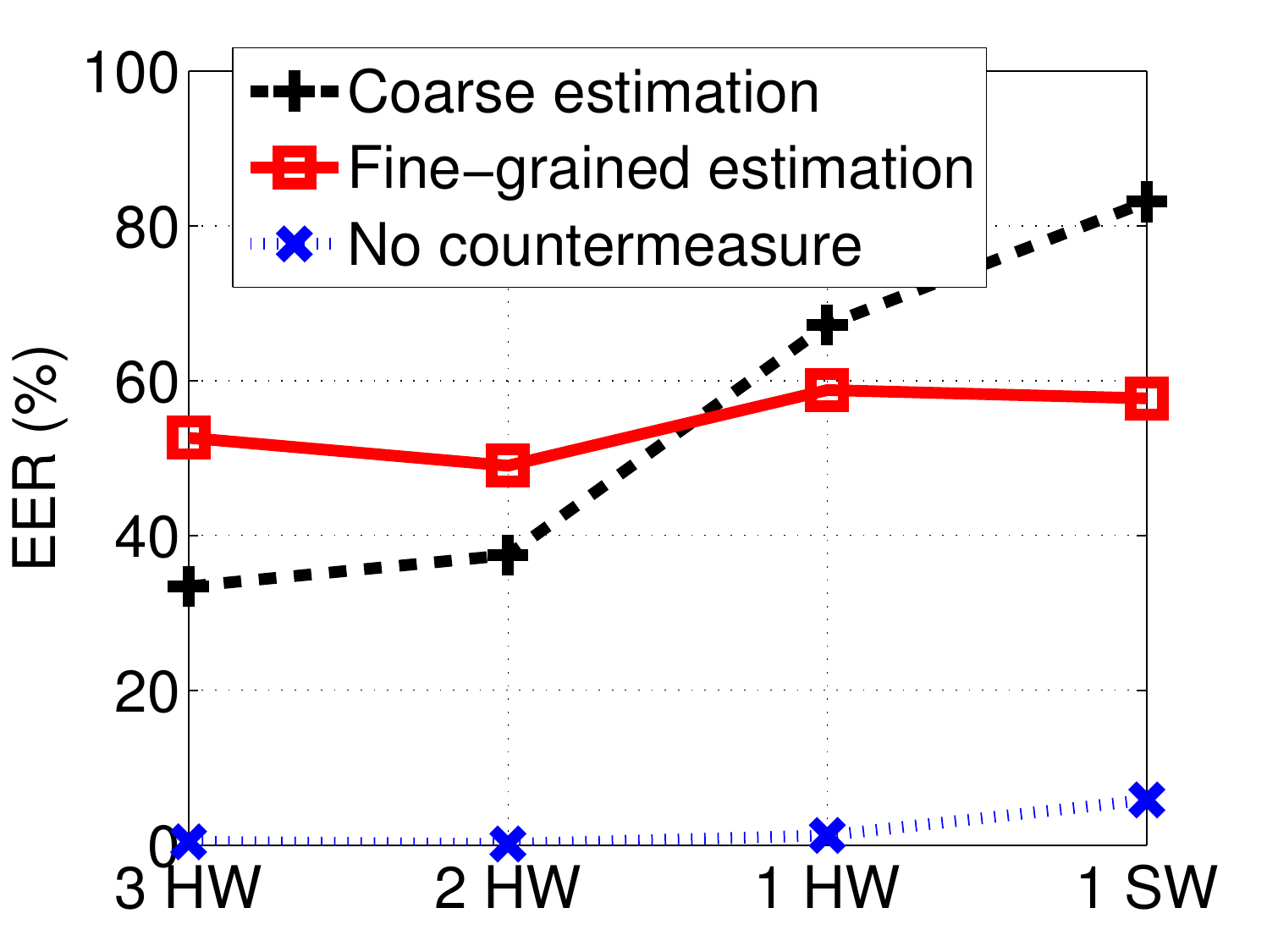}}
  \caption{Impact of the estimation of the delay distribution on the measured EER.}
  \label{fig:eer_countermeasure_test1}
\end{figure}

Similarly, we also argue that lower fingerprinting accuracies can be obtained with our countermeasure if the delay element is equipped with fine-grained delay distributions with respect to the different number of hardware switches that need to be configured in the network. We validate this hypothesis in a separate experiment. Here, we assume the delay element is equipped with best-fit estimates of the distributions of rule installation delays exhibited by both our features with respect to the number of switches in the network, and we measure the corresponding EER witnessed by a remote adversary in our testbed (cf. Figure~\ref{fig:countermeasure_topo}). Our results in Figure~\ref{fig:eer_countermeasure_test1} confirm our hypothesis, and show that when the delay element is equipped with fine-grained information about the distributions of rule installation delays in the network, the EER is closer to 50\%. For example, in this case, the EER increases to almost 40\% using both of our features when the network comprises a software switch, and is almost 47\% when two hardware switches need to be configured. This shows that our countermeasure considerably reduces the distinguishing advantage of a remote adversary, when fine-grained delay distributions are available to the delay element.

\section{Related Work}\label{sec:related}

This paper extends our prior work on fingerprinting SDN networks~\cite{fingerprintingSDN}. The additional contributions are summarized as follows. \emph{(i)} Our evaluation shows that fingerprinting of SDN networks
  with software switches is also feasible. \emph{(ii)} We also investigate the fingerprinting accuracy when the link bandwidth increases to 1\,Gbps. \emph{(iii)}
We conducted further measurements to evaluate the effectiveness of
  fingerprinting SDN networks over a substantial period of
  time. \emph{(iv)} We discuss the implications of our findings on the
  security of SDN networks. \emph{(v)} Finally, we analyze and evaluate our countermeasure.

In the remainder of this section, we discuss related work in the areas of network fingerprinting and SDN security.

\subsection{Network Fingerprinting}

Network fingerprinting has attracted considerable attention in the research community.
Markopoulou \emph{et al.}~\cite{Markopoulou06:WANDelayLoss} show that network delays in backbone networks are
relatively stable and are only marginally affected by network congestion.
Dischinger \emph{et al.}~\cite{Dischinger07:ResidentialNet} show that network features such as
bandwidth and delays mainly depend on
the last-mile hops in residential networks (e.g., due to ISP traffic shaping).
Schulman and Spring~\cite{Schulman11:Pingin} extend this observation by showing that end-to-end delays in residential networks are largely affected by
weather conditions. These findings confirm our observation that the RTT is not a stable feature over time.

Packet-pair dispersion was first proposed to estimate available bandwidth~\cite{Hu03IGI, Strauss03:Spruce,Keshav91:CAFC} and the bottleneck bandwidth of a network path~\cite{Croce08:DSLprobe, Dovrolis04Pathrate, Karame13:SecDisp, Sariou02:SProbe,secure_bandwidth_2009}. Sinha \emph{et al.}~\cite{Sinha06:Fingerprint} observe that the
distribution of packet-pair dispersions can be used to fingerprint the
Internet paths. Karame \emph{et al.}~\cite{Karame13:SecDisp} show that
the packet-pair
dispersion technique is a stable feature which can be used to characterize Internet paths.

\subsection{SDN Security}
A comprehensive survey of security issues in SDN can be found
in~\cite{sdn_survey}.

Shin and Gu~\cite{Shin13:attacking_SDN} briefly hint on the possibility of fingerprinting SDN networks by leveraging timing information of the
exchanged packets; however, in contrast to
our work, their study is not based on a real-world
evaluation and does not provide any metrics to quantify
fingerprinting accuracy.
So-called topology attacks of SDN networks are studied
in~\cite{SPHINX,Hon_etal:ndss2015}. Dhawan~\emph{et al.}~\cite{SPHINX}
also discuss DoS attacks, similar to the ones outlined in
Section~\ref{sec:impl}. They describe an extension of the controller
with a monitoring unit, which detects and reports abnormal behavior.

Shin \emph{et al.}~\cite{rosemary2014} outline a number of vulnerabilities
in current SDN controllers such as Floodlight~\cite{floodlight},
Beacon~\cite{Beacon}, OpenDaylight~\cite{OpenDaylight}, and POX~\cite{POX};
these vulnerabilities allow malicious applications to tamper with the internal data
structures maintained by the SDN controller in order to attack the entire SDN network.
The authors propose a sandbox approach for network applications to deter this misbehavior.
Further mechanisms for securing the control plane are proposed in~\cite{Porras_etal:ndss2015}.
Similarly, AVANT-GUARD~\cite{AVANT-GUARD13} proposes two data plane extensions to enhance the resilience of an SDN network against network flooding attacks
and expedite access to critical data plane activity patterns. We point out that these prior security solutions do not deter fingerprinting
attacks on SDN networks. To the best of our knowledge, our
countermeasure emerges as the only workable solution to alleviate SDN
fingerprinting attacks.  Moreover, we are the first to discuss
information leakage concerning the packet-forwarding logic in SDN.

\section{Conclusion}\label{sec:conclusion}

In this paper, we studied the fingerprinting of SDN networks by a
remote adversary.  For that purpose, we collected measurements from a
large number of hosts located across the globe using a realistic SDN
network.  Our evaluation shows that, by leveraging information from
the RTT and packet-pair dispersion of the exchanged packets,
fingerprinting attacks on SDN networks succeed with overwhelming
probability. Our results also suggest that fingerprinting attacks are
not restricted to active adversaries, but can also be mounted by
passive adversaries that capture a snapshot of the traffic exchanged
with the SDN network.

Based on our results, we presented and evaluated a countermeasure that leverages the
switches' group tables in order to delay the first few packets of
every flow. Our evaluation results show that our countermeasure considerably reduces the ability of an adversary to mount fingerprinting attacks against SDN networks.

\section*{Acknowledgments}

This research was partly performed within the 5G-ENSURE project
(\url{www.5GEnsure.eu}) and received funding from the EU Framework Programme
for Research and Innovation Horizon 2020 under grant agreement
no.~671562.

\bibliographystyle{abbrv}
\balance
\bibliography{references}

\end{document}